\documentclass[9pt,twocolumn,twoside]{pnas-new}

\templatetype{pnasresearcharticle} 
\usepackage[normalem]{ulem}
\usepackage{color}

\newcommand{\supplement}{{\textit{SI Appendix}}}
\usepackage{todonotes}
\newcommand{\figdisp}[1]{Fig. \ref{#1}}

\title{Intertwined spin, charge and pair correlations in the two-dimensional Hubbard model in the thermodynamic limit}

\author[a]{Peizhi Mai}
\author[a]{Seher Karakuzu} 
\author[b]{Giovanni Balduzzi}
\author[c]{Steven Johnston}
\author[a,1]{Thomas A. Maier}

\affil[a]{Computational Sciences and Engineering Division, Oak Ridge National Laboratory, Oak Ridge, TN, 37831-6494, USA.}
\affil[b]{Institute for Theoretical Physics, ETH Zurich, 8093 Zurich, Switzerland.}
\affil[c]{Department of Physics and Astronomy, University of Tennessee, Knoxville, TN 37996-1200, USA.}

\leadauthor{Mai} 

\significancestatement{The high-temperature superconducting cuprates are governed by intertwined striped magnetic and charge orders in addition to superconductivity. Remarkably similar behavior has also been seen in numerical calculations for the Hubbard model describing the copper-oxygen layers in these materials. Finite cluster methods typically find that spin and charge stripe order dominates, while embedded quantum cluster methods, which access the thermodynamic limit, often conclude that superconductivity does. Here, we report the observation of fluctuating spin and charge stripes in an embedded cluster calculation for the Hubbard model. This discovery demonstrates that striped states survive in the thermodynamic limit and allows us to study their influence on the model's superconducting properties, where we find evidence for  pair-density-wave correlations intertwined with the stripe correlations.}

\authorcontributions{DCA Simulations: PM; 
DQMC Simulations: PM; Code Development: PM, SK, GB, SJ, and TAM; Supervision: SJ and TAM; Funding acquisition: SJ and TAM; Writing original draft: PM, SJ, and TAM; Writing review \& editing: All authors.}
\authordeclaration{The authors declare no competing interests. }

\correspondingauthor{\textsuperscript{1}To whom correspondence should be addressed. E-mail:  maierta@ornl.gov}


\begin{abstract}
The high-temperature superconducting cuprates are governed by intertwined spin, charge, and superconducting orders. While various state-of-the-art numerical methods have demonstrated that these phases also manifest themselves in doped Hubbard models, they differ on which is the actual ground state. Finite cluster methods typically indicate that stripe order dominates while embedded quantum cluster methods, which access the thermodynamic limit by treating long-range correlations with a dynamical mean field, conclude that superconductivity does. Here, we report the observation of fluctuating spin and charge stripes in the doped single-band Hubbard model using a quantum Monte Carlo dynamical cluster approximation (DCA) method. By resolving both the fluctuating spin and charge orders using DCA, we demonstrate that they survive in the doped Hubbard model in the thermodynamic limit. This discovery also provides a new opportunity to study the influence of fluctuating stripe correlations on the model's pairing correlations within a unified numerical framework. Using this approach, we also find evidence for pair-density-wave correlations whose strength is correlated with that of the stripes.
\end{abstract}

\dates{This manuscript was compiled on \today}
\doi{\url{www.pnas.org/cgi/doi/10.1073/pnas.XXXXXXXXXX}}

\begin{document}

\maketitle
\thispagestyle{firststyle}
\ifthenelse{\boolean{shortarticle}}{\ifthenelse{\boolean{singlecolumn}}{\abscontentformatted}{\abscontent}}{}

\dropcap{A} common element of strongly correlated materials is the existence of several nearly degenerate states, which compete or cooperate to produce novel phases of matter~\cite{RevModPhys.87.457}. For example, in the high-temperature (high-$T_c$) superconducting cuprates, multiple theoretical studies and experiments point to intertwined orders of spin and charge stripes, charge- and pair-density-waves, and unconventional superconductivity \cite{TranquadaNature1995, RevModPhys.87.457, PDWReview, TranquadaReview, Qin2020, Qin2021, CDWReview}. Understanding the relationships among these orders and how they shape the cuprate phase diagram is a central problem in condensed matter physics.

\begin{figure*}[ht]
    \centering
    \includegraphics[width=1.0\textwidth]{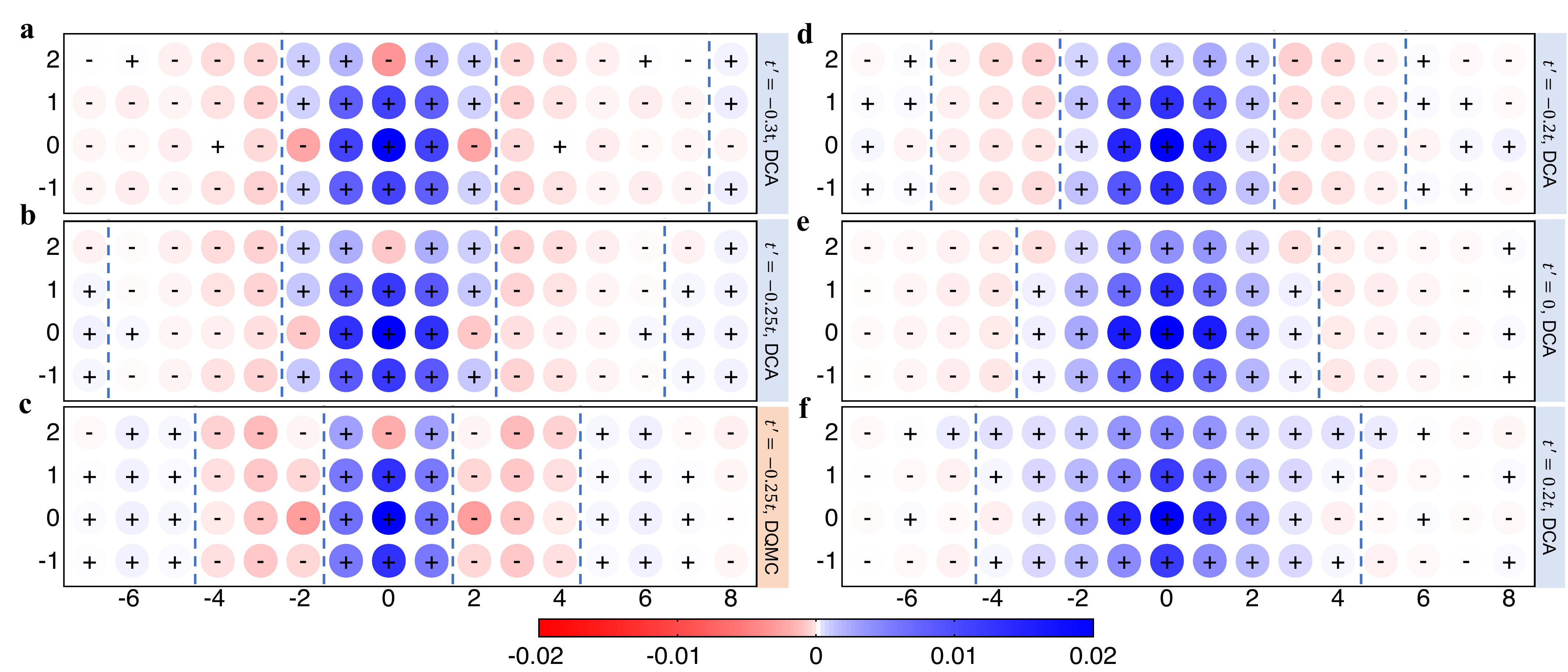}
    \caption{The real-space static staggered spin-spin correlation function of the single-band Hubbard model with $\langle n\rangle=0.8$, obtained from DCA and DQMC simulations. Results are shown for {\bf a} $t^\prime = -0.3t$, DCA; {\bf b} $t^\prime = -0.25t$, DCA; {\bf c} $t^\prime = -0.25t$, DQMC; {\bf d} $t^\prime = -0.2t$, DCA; {\bf e} $t^\prime = 0$, DCA; and {\bf f} $t^\prime = 0.2t$, DCA. The DCA results were obtained using a $16\times4$ cluster embedded in a dynamical mean-field and at an inverse temperature $\beta=6/t$ (Panels {\bf a}-{\bf b} and {\bf d}-{\bf f}). The DQMC results shown in Panel {\bf c} were obtained on a $16\times 4$ cluster with periodic boundary conditions and  $\beta=4.5/t$ and $t^\prime=-0.25t$. Note that here and throughout, we have adopted the same custom color bars used in Refs.~\cite{HuangQuantMat2018} and \cite{HuangScience2017}. This scale provides a finer gradation of small values of the correlation function and improves the overall contrast (see \supplement). 
    }
    \label{fig:spin_stripe}
\end{figure*}

\begin{figure*}[ht]
    \centering
    \includegraphics[width=1.0\textwidth]{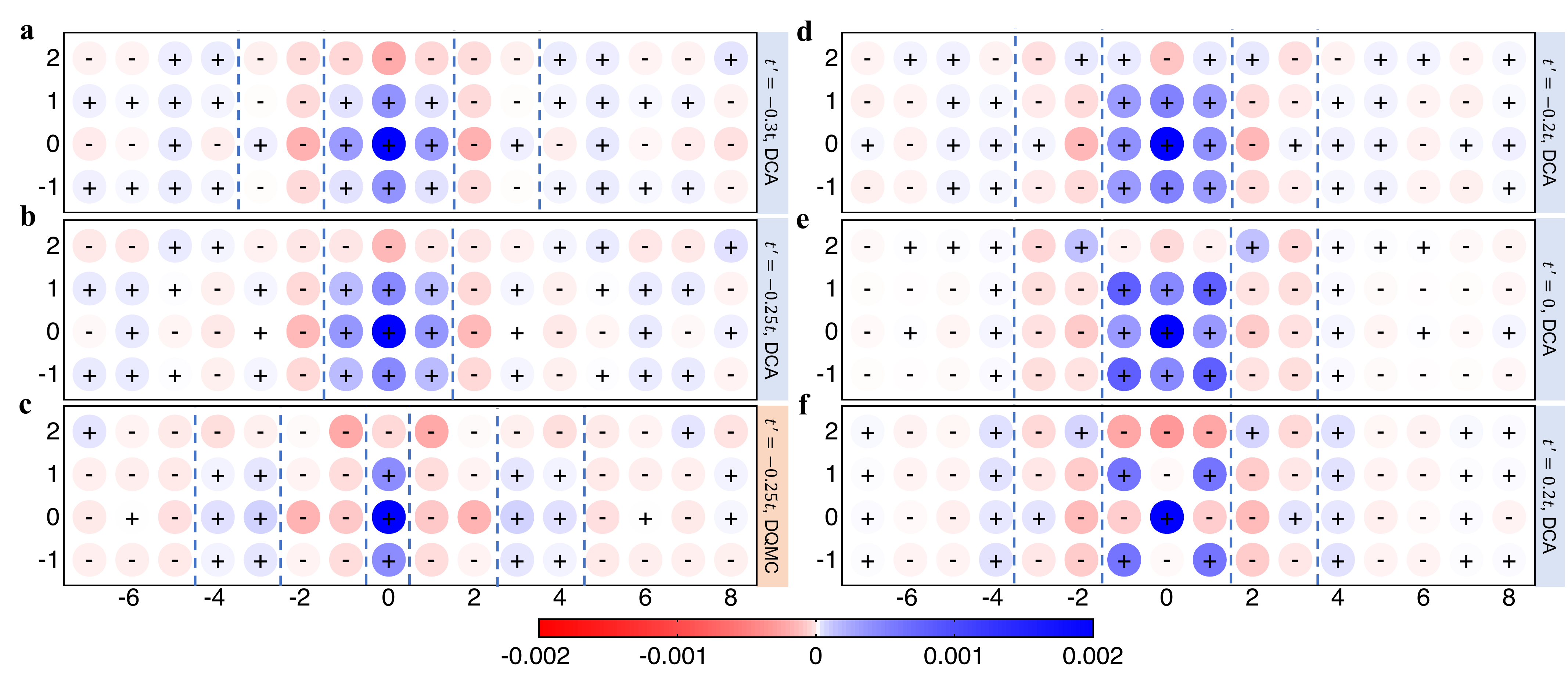}
    \caption{The real-space static density-density correlation function of the single-band Hubbard model, obtained from DCA and DQMC simulations. Results are shown for parameters in a one-to-one correspondence with those shown in Fig.~\ref{fig:spin_stripe}. 
    }
    \label{fig:charge_stripe}
\end{figure*}

\begin{figure*}[ht]
    \centering
    \includegraphics[width=1.0\textwidth]{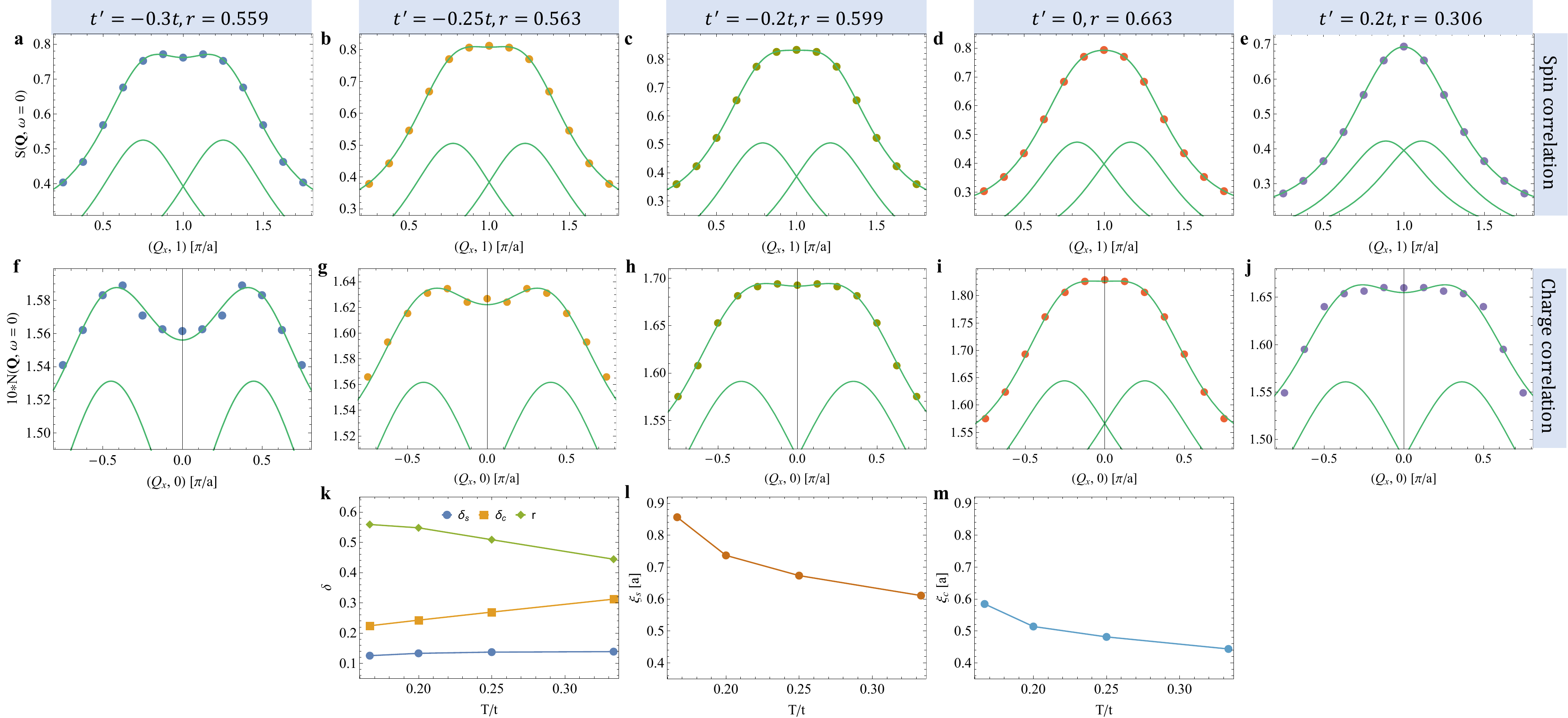}
    \caption{DCA results for the static spin $S({\bf Q},\omega=0)$ (Panels {\bf a-e}) and charge $N({\bf Q},\omega=0)$ (Panels {\bf f-j}) susceptibilities, obtained on a $16\times4$ cluster embedded in a dynamical mean field and with $\langle n\rangle=0.8$. The top row shows the spin susceptibilities along ${\bf Q} = (Q_x,\pi)$ for {\bf a} $t^\prime = -0.3t$, {\bf b} $t^\prime=-0.25t$, {\bf c} $t^\prime=-0.2t$, {\bf d} $t^\prime=0$, and {\bf e} $t^\prime=0.2t$. Each curve is fit with a pair of Lorentzian functions centered at $(2\pi/a)(0.5\pm \delta_s,0.5)$ plus a constant background. The bottom row shows the corresponding charge susceptibilities along ${\bf Q} = (Q_x,0)$. Each curve is fit with a pair of Lorentzian functions centered at $(2\pi/a)(\pm\delta_c,0)$ plus a constant background. All results were obtained for $T = 0.167t$ ($\beta = 6/t$). The ratio of the spin and charge incommensurabilities is given by $r = \delta_s/\delta_c$ and panel {\bf k} shows the temperature evolution of the spin and charge incommensurabilities and their ratio $r$ for $t^\prime=-0.3t$. Panels {\bf l} and {\bf m} show the temperature-dependence of the spin and charge correlation lengths, respectively, for the same $t^\prime$.
    }
    \label{fig:DCA_Stripe_k}
\end{figure*}

\begin{figure}[ht!]
    \centering
    \includegraphics[width=\columnwidth]{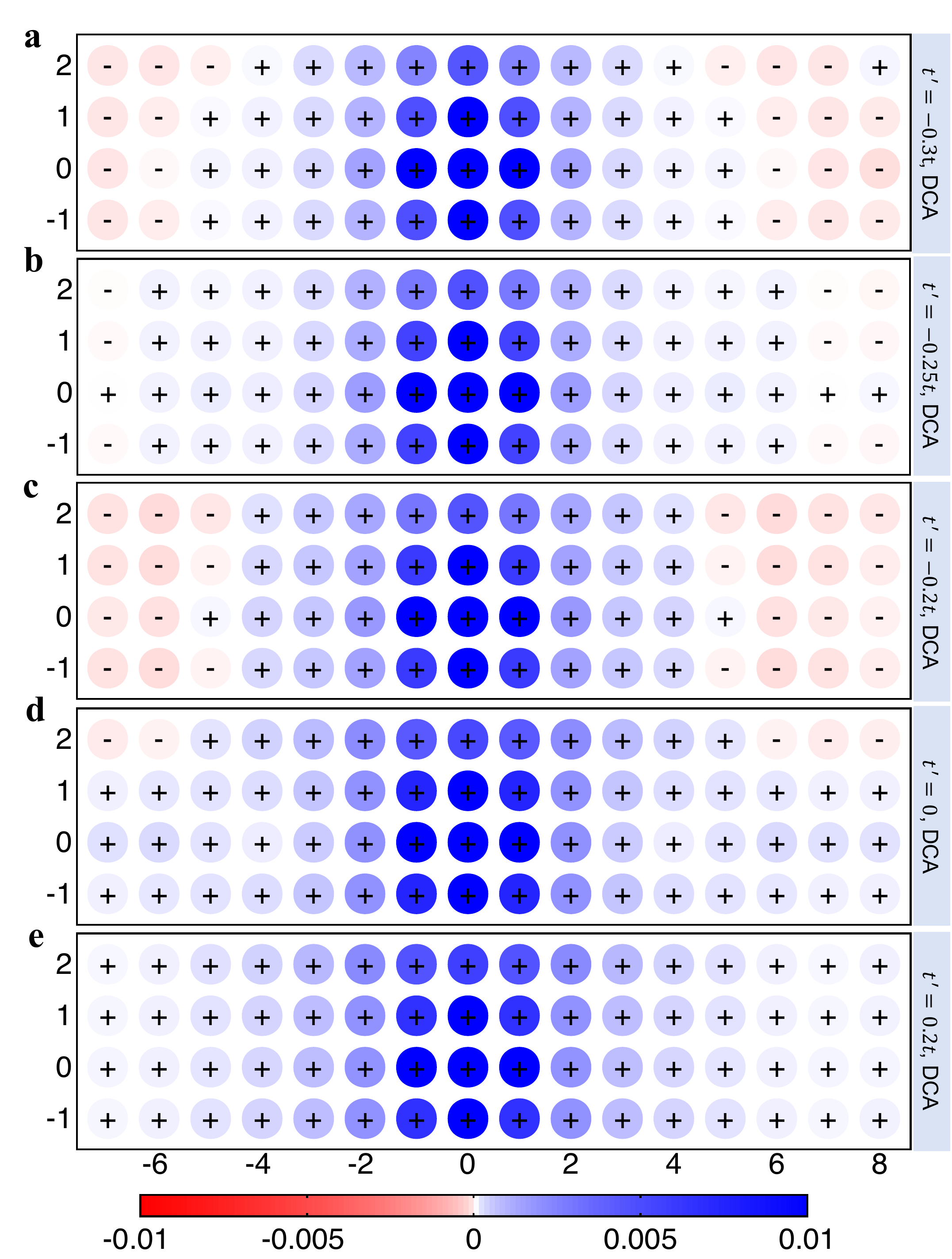}
    \caption{The real-space static $d$-wave pairfield correlation function of the single-band Hubbard model. Results are shown for parameters in one-to-one correspondence with those shown in Fig.~1.}
    \label{fig:Pairs_Fluctuating_Stripes}
\end{figure}

\begin{figure}[ht!]
    \centering
    \includegraphics[width=1.0\columnwidth]{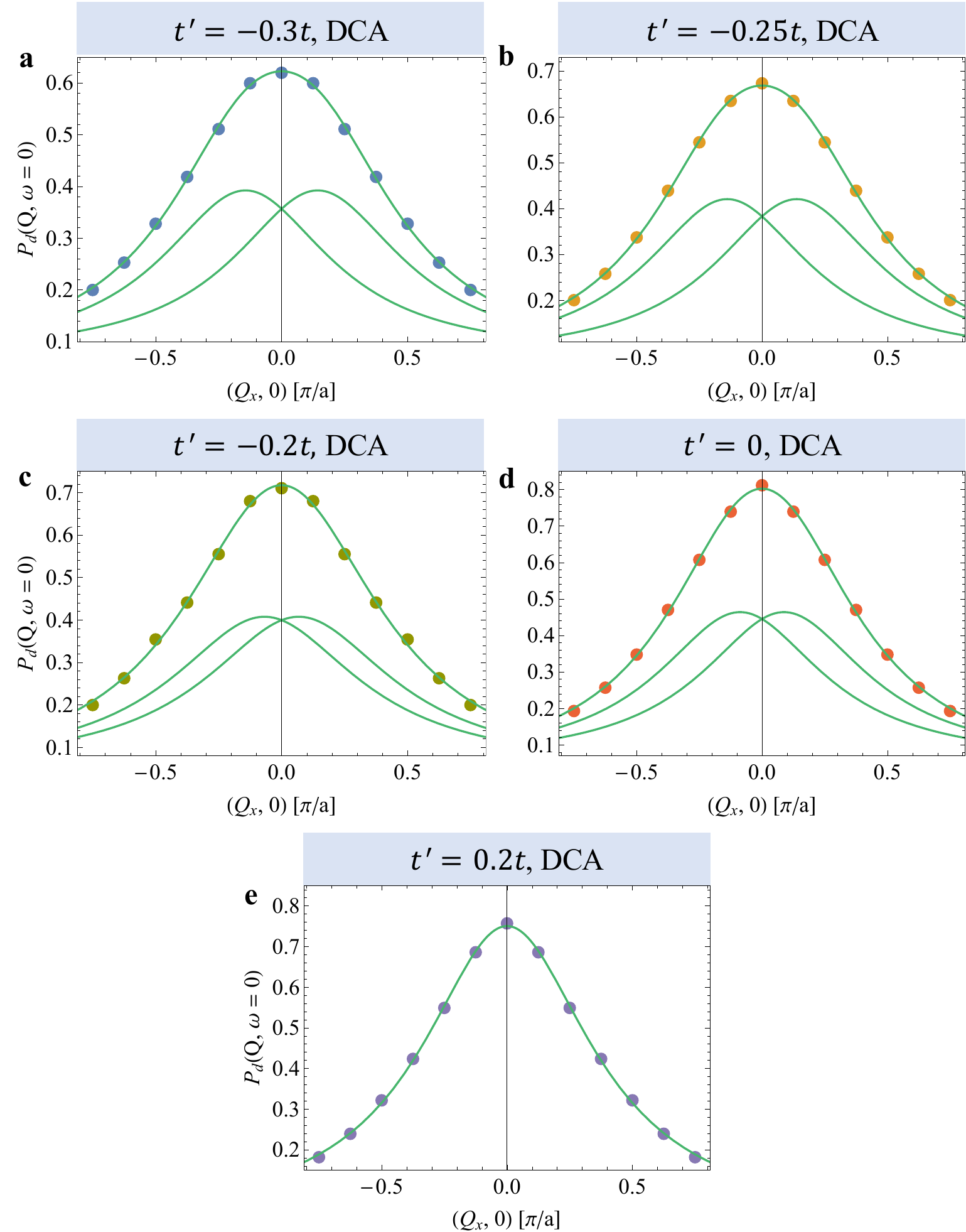}
    \caption{ 
     DCA results for the static pairfield susceptibility $P_d({\bf Q}, \omega=0)$ for $\langle n\rangle=0.8$ and $T = 0.167t$ ($\beta = 6/t$) obtained on a $16\times 4$ cluster embedded in a dynamical mean-field for {\bf a} $t^\prime=-0.3t$, {\bf b} $t^\prime=-0.25t$, {\bf c} $t^\prime=-0.2t$, {\bf d} $t^\prime=0$ and {\bf e} $t^\prime=0.2t$. Each curve is fit with a pair of Lorentzian functions centered at $(\pm\delta_P,0)$. For $t^\prime=0.2t$, the two Lorentzians collapse onto a single peak centered at $(0,0)$. }
    \label{fig:Pairs_momentum}
\end{figure}

Addressing this question using nonpertubative methods remains challenging as even the simplest correlated electron models also contain near-degenerate orders, which can be difficult to discern from one another. For example, state-of-the-art numerical studies have identified a plethora of low-energy states in the single-band Hubbard and $t$-$J$ models that contend for the ground state \cite{ZaanenPRB1989,Machida1989,Kato1990, ZhengScience2017,WhitePRL2003, JiangScience2019, PhysRevResearch.2.033073,PhysRevB.93.035126, HuangQuantMat2018, HuangScience2017, PhysRevB.97.045138, sorella2021phase, CorbozPRL2014,MaierPRL2005, MaierPRL2006, GullEPL2008, SordiPRL2012,Qin2020,Wietek2021,Qin2021}.  Methods like Hartree-Fock mean-field theory \cite{ZaanenPRB1989,Machida1989,Kato1990}, density matrix renormalization group (DMRG) \cite{WhitePRL2003, ZhengScience2017,JiangScience2019,PhysRevResearch.2.033073,Qin2020}, density matrix embedding theory \cite{PhysRevB.93.035126, ZhengScience2017},  variational Monte Carlo~\cite{PhysRevB.97.045138}, auxiliary field quantum Monte Carlo \cite{ZhengScience2017, sorella2021phase,Qin2020,Qin2021}, and infinite projected entangled-pair states \cite{CorbozPRL2014,ZhengScience2017} tend to find static stripe order, i.e. unidirectional spin and charge density waves, as the ground state, and recent determinant quantum Monte Carlo (DQMC) calculations \cite{HuangQuantMat2018, HuangScience2017} and minimally-entangled typical thermal states (METTS) \cite{Wietek2021} have found evidence for spin stripe correlations at finite temperatures. In contrast, quantum cluster methods like cellular dynamical mean-field theory \cite{Foley2019,Kotliar2001,SordiPRL2012} and the dynamical cluster approximation (DCA)~\cite{JarrellPRB2001, MaierRMP2005}, which, unlike finite size cluster techniques, directly access the thermodynamic limit, typically find superconducting solutions with a $d$-wave symmetry \cite{MaierPRL2005, MaierPRL2006, GullEPL2008, SordiPRL2012}. While approximate calculations of this sort with large unit cells but correlations restricted to single sites or small clusters have found evidence of stripes \cite{Robert2014,Fleck2000,Vanhala2018,Dash2021}, more reliable calculations with clusters large enough to accommodate the stripe periodicity have yet to find any indication of stripe-like solutions. 
It is crucial that we understand and resolve these differences to properly identify the properties of the Hubbard model in the thermodynamic limit. Moreover, this dichotomy has made it difficult to understand the relationships between the relevant orders since each method has its approximations, which introduce systematic errors that can bias towards particular solutions. To overcome this issue, it is desirable to identify a single framework capable of identifying the relevant states to avoid compounding systematic biases. 

Here, we demonstrate that quantum Monte Carlo DCA \cite{JarrellPRB2001} methods can detect short-range stripe correlations in the two-dimensional single-band Hubbard model. 
The QMC-based impurity solver captures the intra-cluster correlations exactly, while longer-range correlations are treated in a mean-field that approximates the infinite system. The observation of fluctuating stripes with this method therefore provides crucial confirmation that such correlations persist in the thermodynamic limit. It also fixes the discrepancy between finite-cluster and quantum cluster methods. Our results allow us to examine the influence of fluctuating stripe correlations on $d$-wave pairing in the single-band Hubbard model using a unified framework, provided the clusters are large enough to accommodate the relevant periodicity such that the spatial modulations are not averaged out by the DCA mean field.  
Using this unified framework, we find evidence for short-range pair-density-wave (PDW) correlations, whose strength is correlated with the strength of the spin and charge stripe correlations.

\section*{Model}
We consider the two-dimensional single-band Hubbard Hamiltonian defined on a rectangular 
$N=N_x\times N_y$ lattice
\begin{equation}\label{Eq:Hubbard}
    H= -\sum_{{\bf i},{\bf j},\sigma} t^{\phantom\dagger}_{{\bf i}{\bf j}}\left(c^\dagger_{{\bf i}\sigma}c^{\phantom\dagger}_{{\bf j}\sigma} + \mathrm{h.c.}\right)  
    -\mu\sum_{{\bf i},\sigma} n_{{\bf i}\sigma} + U\sum_{{\bf i}}n_{{\bf i}\uparrow}n_{{\bf i}\downarrow}. 
\end{equation}
Here, $c^\dagger_{{\bf i}\sigma}$ ($c^{\phantom\dagger}_{{\bf i}\sigma}$) creates (annihilates) 
a spin-$\sigma$ ($=\uparrow,\downarrow$) electron on site ${\bf i}$; $n^{\phantom\dagger}_{{\bf i}\sigma} = c^\dagger_{{\bf i},\sigma}c^{\phantom\dagger}_{{\bf i}\sigma}$ is the number operator; $t_{{\bf i}{\bf j}}$ is the hopping integral between sites ${\bf i}$ and ${\bf j}$; 
$\mu$ is the chemical potential; and $U$ is the on-site Hubbard repulsion. Throughout, we restrict $t_{{\bf i}{\bf j}}$ to nearest- ($t$) and next-nearest-neighbor ($t^\prime$) hopping only and set $U=6t$ to facilitate comparisons to Ref.~\cite{HuangQuantMat2018}. 
We then solved Eq.~(\ref{Eq:Hubbard}) using DCA~\cite{JarrellPRB2001} and with a continuous-time QMC impurity solver \cite{GullEPL2008}, as implemented in the DCA++ code~\cite{UrsCompPhysComm2020}, and complementary DQMC calculations~\cite{BlankenbeclerPRD1981, WhitePRB1989}. 

Previous DQMC calculations \cite{HuangScience2017, HuangQuantMat2018} have demonstrated that the high-temperature spin-stripe correlations in the single- and multi-band Hubbard model are fluctuating in nature, where short-range spatial correlations appear over several unit cells and are fluctuating in time. 
In general, static ($\omega=0$) correlation functions, which integrate over the imaginary time dynamics, are expected to be more sensitive to fluctuating short-range order \cite{KivelsonRMP2003} compared to the corresponding equal-time ($\tau = 0$) correlation functions \cite{HuangQuantMat2018}. We have found that this is indeed the case for the spin, charge, and PDW-like correlations reported here (see \supplement), supporting the interpretation that the observed stripe correlations are fluctuating in nature. 

To study the fluctuating spin stripes, we measured the static staggered spin-spin correlation function $S^\mathrm{stag}({\bf r},\omega=0)=(-1)^{r_x+r_y}\tfrac{1}{N}\int_0^\beta \sum_{\bf i} \langle \hat{S}^z_{\bf i+r}(\tau)~\hat{S}^z_{\bf i}(0)\rangle d\tau$, where ${\bf r} = a(r_x,r_y)$ is the position of each atom on the square lattice with lattice constant $a$ and $\hat{S}_{\bf i}^z =  \frac{1}{2}\left(c^\dagger_{{\bf i},\uparrow}c^{\phantom\dagger}_{{\bf i},\uparrow} - c^\dagger_{{\bf i},\downarrow}c^{\phantom\dagger}_{{\bf i},\downarrow}\right)$ is the $z$-component of the local spin operator at site ${\bf i}$. The fluctuating charge stripe correlations are assessed by measuring the static density-density correlation function 
$N({\bf r},\omega=0) = \tfrac{1}{N} \int_0^\beta \sum_{\bf i} \left(\langle n_{{\bf i}+{\bf r}}(\tau)~n_{\bf i}(0)\rangle - \langle n_{{\bf i}+{\bf r}}(\tau)\rangle\langle n_{\bf i}(0)\rangle\right) d\tau$, where $n_{\bf i} = \sum_\sigma n_{{\bf i},\sigma}$ is the local density operator. The pairing tendencies are accessed by measuring the static pairing correlation function in the $d$-wave channel $P_d({\bf r},\omega=0) = \frac{1}{N} \int_0^\beta\sum_{\bf i}\langle \Delta^{\phantom\dagger}_{{\bf i}+{\bf r}}(\tau)~\Delta^\dagger_{\bf i}(0) \rangle d\tau$, where $\Delta_{\bf i}=c_{{\bf i},\uparrow}(c_{{\bf i}+\hat{x},\downarrow}+c_{{\bf i}-\hat{x},\downarrow}-c_{{\bf i}+\hat{y},\downarrow}-c_{{\bf i}-\hat{y},\downarrow})-c_{{\bf i},\downarrow}(c_{{\bf i}+\hat{x},\uparrow}+c_{{\bf i}-\hat{x},\uparrow}-c_{{\bf i}+\hat{y},\uparrow}-c_{{\bf i}-\hat{y},\uparrow})$ destroys a singlet pair of electrons with $d$-wave symmetry. We also determined the structure of the pairing interaction by explicitly solving the Bethe-Salpeter equation (BSE) in the particle-particle singlet channel to obtain its leading eigenvalues and eigenvectors \cite{MaierPRL2006, Mai2021}. Due to the large cluster sizes and the self-consistency loop, our DCA calculations are significantly more expensive than the corresponding DQMC calculations. For this reason, we focus on an average density $\langle n\rangle=0.8$, where we have observed strong stripe correlations.

\section*{Results and Discussion}

Figure \ref{fig:spin_stripe} plots $S^\mathrm{stag}({\bf r})$ for several values of the next-nearest neighbor hopping $t^\prime$, where we see clear evidence for spin stripe correlations in our DCA calculations (Panels a, b, d-f). Here, we employ large $16\times 4$ clusters embedded in the DCA self-consistent mean-field, where we access temperatures as low as $T=0.167t$ (inverse temperature $\beta = 6/t$). Since the staggered spin-spin correlation function imposes a sign flip on every other site, the positive blue regions in the middle of each panel represent short-range antiferromagnetic (AFM) correlations. In contrast, the negative red regions represent AFM regions but with a $\pi$ phase shift. As $t^\prime/t$ decreases from positive to negative, red negative regions form more prominently on both sides of the central blue region, signaling the formation and growth of AFM stripe fluctuations, similar to those observed in finite size DQMC calculations \cite{HuangScience2017, HuangQuantMat2018}. In general, we find that the boundary between the red and blue regions mixes, suggesting that the stripes are incommensurate. We have observed similar patterns for different cluster sizes and geometries, including $8\times8$, $8\times6$ and $8\times 4$ clusters, and in the spin $xx$ correlations (see \supplement).

For comparison, \figdisp{fig:spin_stripe}{\bf c} shows $S^\mathrm{stag}({\bf r},\omega=0)$ obtained from a DQMC calculation at $T = 0.22t$ ($\beta = 4.5/t$), $t^\prime=-0.25t$. The DQMC results are consistent with the corresponding DCA results (Fig.~\ref{fig:spin_stripe}b). (A systematic comparison between DCA and an earlier DQMC study \cite{HuangQuantMat2018} focusing on the real-space equal-time spin-spin correlation function at the density $n=0.875$ is given in \supplement.) Since DQMC treats the system exactly on an extended but finite cluster, one must perform a finite-size scaling analysis to access the thermodynamic limit. On the other hand, DCA accesses the thermodynamic limit by embedding its clusters in a dynamical mean field that approximates the rest of the system. Comparing  Figs.~\ref{fig:spin_stripe}b and ~\ref{fig:spin_stripe}c, we find that DCA predicts weaker stripe correlations compared to DQMC for the same $t^\prime$, despite the lower temperature. This observation may help explain why stripes have previously gone unobserved in quantum cluster approaches employing extended clusters. The origin of the reduced correlations is unclear at this time. One possibility is that the correlations observed by DQMC would weaken as the cluster size increases. Another is that the mean field reduces the effective correlations in the DCA treatment of the problem or tends to favor uniform states and restore $C_4$ symmetry in the cluster. Nevertheless, the observation of stripes with DCA provides crucial evidence that they persist in the thermodynamic limit. 

Although we observe similar fluctuating spin stripes in both zero-frequency (Fig.~\ref{fig:spin_stripe}) and equal-time (\supplement) spin-spin correlation functions, the zero-frequency (Fig.~\ref{fig:charge_stripe}) and equal-time (\supplement) density-density correlation functions show qualitatively different behaviors. The density-density correlations have much stronger imaginary time dependence. A detailed imaginary time analysis can be found in \supplement. As a result, the fluctuating charge stripe pattern is only observed in the static ($\omega=0$) correlation functions. Fig.~\ref{fig:charge_stripe} plots $N({\bf r},\omega=0)$ for the same $16\times 4$ DCA and DQMC simulations shown in Fig.~\ref{fig:spin_stripe}. We observe the central blue region surrounded by two red regions on both sides, signaling short-range charge stripe fluctuations but with a shorter period. As $t^\prime/t$ increases from negative to positive, unlike the spin case where the blue region extends to weaken the stripe, the charge blue region does not extend. In the electron-doped case ($t^\prime=0.2t$), however, the central region is dominated by a staggered $(\pi,\pi)$ CDW-like correlation, with the sub-dominant stripe-like pattern laying on both sides. As a comparison, the DQMC result in Fig.~\ref{fig:charge_stripe}c shows a stronger stripe pattern with a shorter period than the corresponding DCA result in Fig.~\ref{fig:charge_stripe}b.


The presence of the spin and charge stripes is more readily observed by examining the dynamical spin $S({\bf Q},\omega)$ and charge $N({\bf Q},\omega)$ susceptibilities \cite{KivelsonRMP2003}, which measure the collective fluctuations. Here, we consider the static limit $(\omega=0)$, which can be obtained by Fourier transforming the corresponding static real-space correlation functions. \figdisp{fig:DCA_Stripe_k} summarizes $S({\bf Q},\omega=0)$ along ${\bf Q}=(Q_x,\pi)$ (top row) and $N({\bf Q},\omega=0)$ along ${\bf Q} = (Q_x,0)$ (bottom row) for different values of the next-nearest-neighbor hopping $t^\prime$. When the spin stripe correlations are strong, they should manifest as incommensurate peaks in $S({\bf Q},0)$ centered at $(2\pi/a)(0.5\pm \delta_s,0.5)$ while the charge stripes should manifest as incommensurate peaks in $N({\bf Q},0)$ centered at $(2\pi/a)(\pm\delta_c,0)$, where $\delta_c = 2\delta_s$. To check this, we fit the spectra with pairs of Lorentzian functions (plus a constant background) and extracted the corresponding values of $\delta_{s,c}$. In all cases, the susceptibilities are well represented by the fits, and the resulting ratio $r=\delta_s/\delta_c$ is given in the top panels of Fig.~\ref{fig:DCA_Stripe_k}, where we find $0.66 > r > 0.3$. Especially in the cases with $t^\prime<0$, these values are close to the expected $r=0.5$ result, for which the periodicity of the charge fluctuations is twice that of the spin fluctuations.

To examine how the spin and charge stripes form, we also extracted the values of $\delta_c$, $\delta_s$, and $r=\delta_s/\delta_c$ as a function of temperature for $t^\prime = -0.3t$, as shown in \figdisp{fig:DCA_Stripe_k}{\bf k}. (The corresponding susceptibility data is provided in \supplement.) We find that $\delta_s$ remains relatively fixed as a function of temperature, while $\delta_c$ appears to lock in to its value of $\delta_c\approx 2\delta_s$ as the temperature is lowered.  From Figs.~\ref{fig:spin_stripe}-\ref{fig:DCA_Stripe_k}, we can see that the incommensurability $\delta_s$ (or period) of the fluctuating spin stripes depends weakly on temperature but strongly on $t^\prime$, while the incommensurability $\delta_c$ (or period) of the fluctuating charge stripes depends strongly on temperature but weakly on $t^\prime$. \figdisp{fig:DCA_Stripe_k}l and \figdisp{fig:DCA_Stripe_k}m show the correlation lengths $\xi_s$ and $\xi_c$ for $t^\prime=-0.3t$, for the fluctuating spin and charge stripes, respectively. These values are determined from the inverse of the half-width-half-maximum of the Lorentzian fits, but we obtain very similar estimates from exponential fits of the real-space correlations. We find that both the spin and charge correlation lengths extracted this way are relatively short at the temperatures we can access but grow with decreasing temperature. We also find that the charge correlation length is substantially shorter than the spin correlation length. These results suggest that in the hole-doped case, the charge stripes emerge at lower energy scales than spin stripes do in the Hubbard model, and that the periodicity of the former locks into the value set by the latter. This finding is counter to the idea that the charge order forms prior to the spin order~\cite{Zachar1998,Berg2007}. In the electron-doped case, the comparison between Figs.~\ref{fig:spin_stripe}{\bf f} and~\ref{fig:charge_stripe}{\bf f} shows that the charge stripe appears at higher energy scale instead.

Previous DCA studies have observed a finite-temperature transition to the $d$-wave superconducting state~\cite{MaierPRL2005, Mai2021}. Now that DCA also finds evidence for fluctuating stripes, both in the spin and charge sector, it is natural to ask how they affect the formation of Cooper pairs. To answer this question, we examine the static $d$-wave pairfield susceptibility $P_d({\bf r},\omega=0)$ calculated with DCA on a 16$\times$4 cluster as shown in Fig.~\ref{fig:Pairs_Fluctuating_Stripes}. Interestingly, as $t^\prime$ is reduced and varied to more negative values, the static pairfield correlations develop a modulated striped pattern with a sign change ($\pi$ phase shift) suggestive of a pair-density-wave \cite{PDWReview}. This trend is similar to that found for the spin stripe fluctuations in Fig.~\ref{fig:spin_stripe}, except that for $t^\prime=-0.25t$ the modulation is less visible possibly due to a change in the Fermi surface topology from electron- to hole-like \cite{Wu2018}.  Fig.~\ref{fig:Pairs_momentum} shows the corresponding plots of the Fourier-transformed static $d$-wave pairfield susceptibility $P_d({\bf Q}, \omega=0)$ fitted with a pair of Lorentzian functions. For $t^\prime=0.2t$, we observe a single peak at ${\bf Q}=(0,0)$. But as $t^\prime$ is lowered and varied to more negative values, one sees that the peak flattens out. In this case, the best fit is obtained with two separate Lorentzians centered at $(2\pi/a)(\pm \delta_P,0)$, consistent with the periodic pair-density-wave like modulation observed in Fig.~\ref{fig:Pairs_Fluctuating_Stripes}.

These observations suggest that striped modulations in the pairfield correlations develop together with the spin and charge stripe correlations. At the temperatures we have studied,  it is clear from Fig.~\ref{fig:Pairs_momentum} that their signature in the momentum structure of the pairfield susceptibility is not as strong as that of the spin and charge stripes. Nevertheless, the striped modulation is clearly visible in the real space structure in Fig.~\ref{fig:Pairs_Fluctuating_Stripes}.  Whether the PDW-like modulations become stronger at lower temperatures and potentially lead to a superconducting instability to a PDW state, however, remains an open question that we are unable to address because of the Fermion sign problem.


\section*{Conclusion}
Our results demonstrate that fluctuating spin and charge stripe orders are a property of the doped single-band Hubbard model in the thermodynamic limit. Moreover, by accessing these phases using an embedded cluster technique, we are able to examine the ways in which the stripe fluctuations couple to superconducting correlations in the model.  Concomitant with the spin stripes, we find that the $d$-wave pairing correlations develop a similar periodic stripe modulation, indicative of a pair-density wave.


\section*{Materials and methods}
\subsection*{Dynamical cluster approximation} 
To study the single-band Hubbard model, we use the dynamical cluster approximation (DCA) \cite{JarrellPRB2001,MaierRMP2005,MaierPRL2005,UrsCompPhysComm2020}. Complete details of the DCA algorithm can be found in Ref.~\cite{MaierRMP2005}. The DCA coarse-grains momentum space to represent the bulk lattice in the thermodynamic limit by a finite size cluster that contains $N_c$ sites and is embedded in a self-consistent mean-field. This mean-field represents the remaining degrees of freedom and is determined self-consistently from the solution of the cluster problem.  

With the assumption of short-ranged correlations, the self-energy $\Sigma({\bf k},i \omega_n)$ is well approximated by the cluster self-energy $\Sigma({\bf K},i \omega_n)$, where ${\bf K}$ are the cluster momenta. The coarse-grained single-particle Green's function
\begin{equation}
\begin{split}
\bar{G}({\bf K},i\omega_n)&=\frac{N_c}{N}\sum_{\bf{k}^\prime}G({\bf K}+{\bf k}^\prime,i\omega_n)
\\&=\frac{N_c}{N}\sum_{\bf{k'}}\frac{1}{i\omega_n+\mu-\varepsilon({\bf K}+{\bf k'})-\Sigma({\bf K},i \omega_n)},
\end{split}
\end{equation}
 is then obtained by averaging the lattice Green's function $G({\bf k},i\omega_n)$ over the $N/N_c$ momenta $\bf{k}^\prime$ in a square patch about the cluster momentum ${\bf K}$ that has an area of size $1/N_c$ of that of the first Brillouin zone. This reduces the bulk problem to a finite size cluster, which we solve using the continuous-time, auxiliary-field, quantum Monte-Carlo algorithm (CT-AUX)~\cite{GullEPL2008}. 
 
In our DCA++ simulations, the expansion order of the CT-AUX QMC is typically in the range of 100 - 3000, depending on temperature and $t^\prime$. Depending on the average fermion sign for a given parameter set, we have obtained 10 million - 2 billion samples for the correlation functions. Usually 6 - 8 iterations were needed to obtain good convergence for the DCA mean-field. 
 
\subsection*{Determinant quantum Monte Carlo}
We also perform DQMC\cite{BlankenbeclerPRD1981, WhitePRB1989} simulation on the single-band Hubbard model to obtain the spin, charge and pairing correlation functions for finite-size clusters. We divide the imaginary time interval $[0,\beta]$ into $L$ discrete steps with step size fixed at $\Delta\tau=0.1$, and rewrite the partition function using the Trotter formula neglecting terms of order $\mathcal{O}(\Delta\tau^2)$. 

We perform equal-time measurements every other full space-time sweep and unequal-time measurements every 4th sweep. We use 5000 independently seeded Markov chains and for each chain, we use 50,000 warmup sweeps and 400,000 measurement sweeps. This large amount of data set leads to reliable statistics despite the severe Fermion sign problem. 

\acknow{The authors would like to thank T. P. Devereaux, E. Huang, B. Moritz, and D. J. Scalapino for useful discussions. This work was supported by the Scientific Discovery through Advanced Computing (SciDAC) program funded by the U.S. Department of Energy, Office of Science, Advanced Scientific Computing Research and Basic Energy Sciences, Division of Materials Sciences and Engineering. This research used resources of the Oak Ridge Leadership Computing Facility, which is a DOE Office of Science User Facility supported under Contract DE-AC05-00OR22725. A part of the analysis of the results performed by T. A. M. was supported by the U.S. Department of Energy, Office of Basic Energy Sciences, Materials Sciences and Engineering Division. This manuscript has been authored by UT-Battelle, LLC under Contract No. DE-AC05-00OR22725 with the U.S. Department of Energy. The United States Government retains and the publisher, by accepting the article for publication, acknowledges that the United States Government retains a non-exclusive, paid-up, irrevocable, world-wide license to publish or reproduce the published form of this manuscript, or allow others to do so, for United States Government purposes. The Department of Energy will provide public access to these results of federally sponsored research in accordance with the DOE Public Access Plan (http://energy.gov/downloads/doe-public-access-plan).\\}

\showacknow{} 

\subsection*{References}

\bibliography{refbib}

\end{document}



\maketitle

\SItext
\section{The spin-$zz$ and $xx$ stripe correlations}
\begin{figure}[h!]
    \centering
    \includegraphics[width=\textwidth]{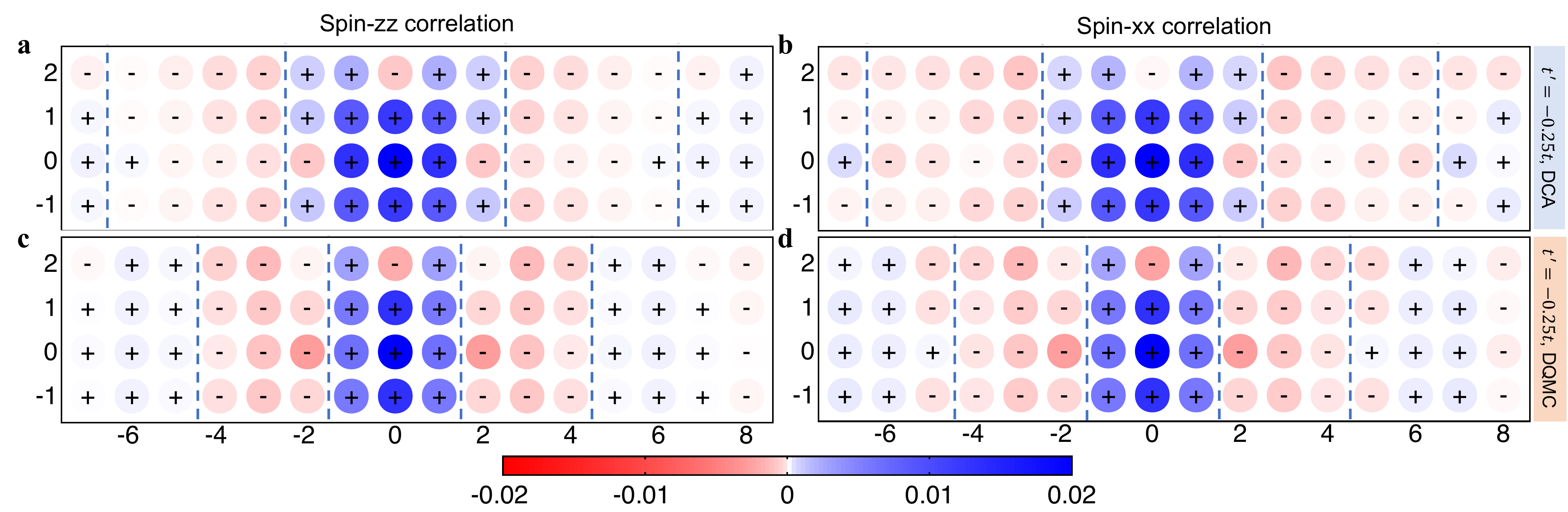}
    \caption{The real-space static staggered spin $zz$ and $xx$ correlation functions of the single-band Hubbard model with $\langle n\rangle=0.8, t^\prime=-0.25t, U=6t$, obtained from DCA ({\bf a}, {\bf b}) and DQMC ({\bf c}, {\bf d}) simulations. The DCA calculation was conducted at $T=0.167t$ ($\beta=6/t$), while the DQMC was at $T=0.22t$ ($\beta=4.5/t$). The $xx$ and $zz$ correlations show consistently the same spin stripe within error bar for each method.}
    \label{fig:spin_zzNxx}
\end{figure}

\figdisp{fig:spin_zzNxx} shows the static staggered spin $zz$ and $xx$ correlation functions from DCA and DQMC simulations. In each method, the $xx$ and $zz$ correlations show essentially the same stripe pattern within error bars. Examining the statistical errors, we find that the relative size of the error bars for the $zz$ and $xx$ correlation functions depends on whether we're examining the $\tau = 0$ or the $\omega = 0$ correlations. Specifically, the $zz$ correlation function has smaller error bars than the $xx$ when we examine the equal-time $\tau = 0$ correlations, while this trend is reversed in the $\omega = 0$ correlation functions. 

\section{The spin and charge stripe correlations for different cluster sizes}
\begin{figure}[h!]
    \centering
    \includegraphics[width=0.73\textwidth]{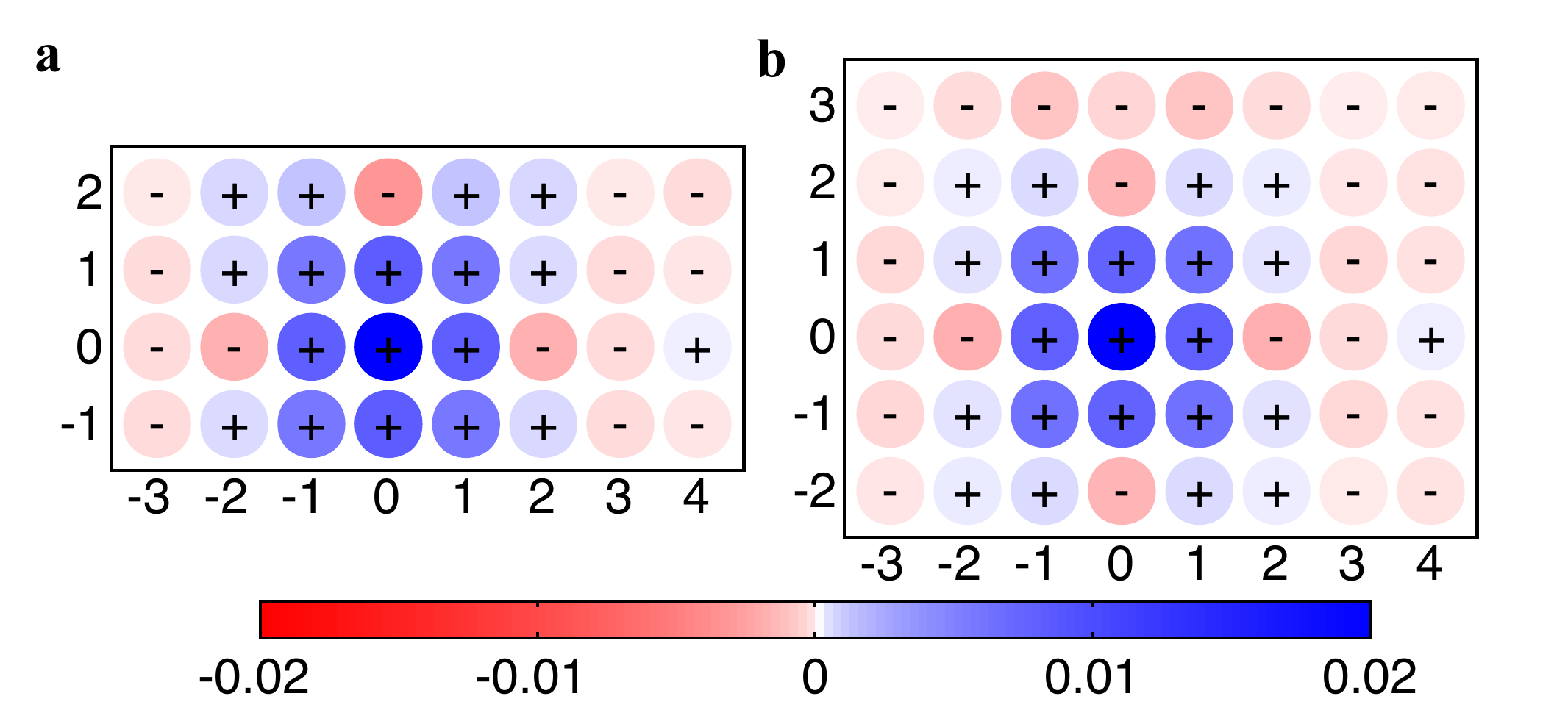}
    \caption{The real-space static staggered spin-spin correlation function of the singleband Hubbard model calculated on an $8\times4$ cluster ({\bf a}) and an $8\times6$ cluster ({\bf b}) at $t^\prime=-0.3t$, $\langle n \rangle = 0.8$, and $T=0.2t$ ($\beta = 5/t$). Both clusters show a spin stripe pattern.}
    \label{fig:spin_stripe_diff_cluster}
\end{figure}

Figure~1 of the main text shows the spin stripe correlation patterns calculated on $16\times4$ clusters embedded in the DCA mean field. \figdisp{fig:spin_stripe_diff_cluster} presents similar results obtained on smaller $8\times4$ and $8\times6$ clusters, which both show indications of spin stripe correlations. For example, the correlations observed in the $8\times4$ case resemble the ones found in the corresponding region of the $16\times4$ case (Fig.~1a). Similarly, the correlations observed in the $8\times6$ case are similar to the corresponding correlations found in the $8\times8$ case shown in Fig.~4a.

\begin{figure}[h!]
    \centering
    \includegraphics[width=0.6\textwidth]{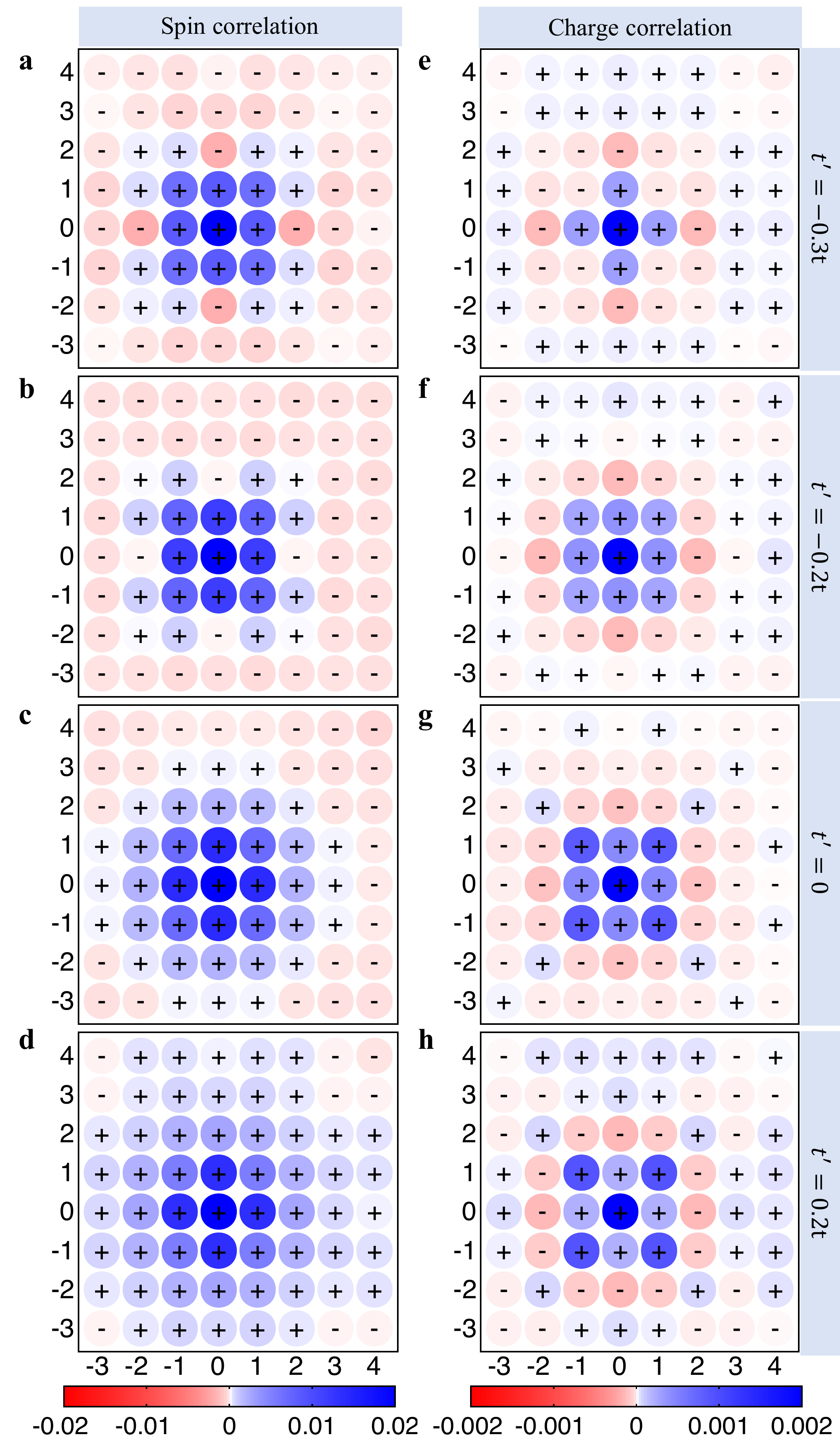}
    \caption{DCA results for the real-space static correlation functions, 
    obtained on $8\times 8$ clusters at temperature $T = 0.2t$ ($\beta = 5/t$), $\langle n\rangle=0.8$,  
    and $t^\prime = -0.3t$ (first row), $t^\prime = -0.2t$ (second row), $t^\prime = 0$ (third row), and $t^\prime = 0.2t$ (fourth row). 
    Panels ({\bf a}-{\bf d}) show results for the staggered spin-spin correlation function. Panels ({\bf e}-{\bf h}) show the corresponding real-space density-density correlation function.}
    \label{fig:8by8_static}
\end{figure}

We also present the DCA results for the static spin and charge correlation functions for a square $N = 8\times8$ cluster in \figdisp{fig:8by8_static}. Since the $8\times 8$ cluster no longer breaks $C_4$ symmetry, both the spin and charge stripe correlations now appear as superimposed vertical and horizontal stripes. As with Figs.~1 and 2 in the main text for the $16\times4$ cluster, their strength increases as $t^\prime$ becomes more negative.

\newpage
\section{The equal-time correlation function}
\begin{figure*}[ht]
    \centering
    \includegraphics[width=1.0\textwidth]{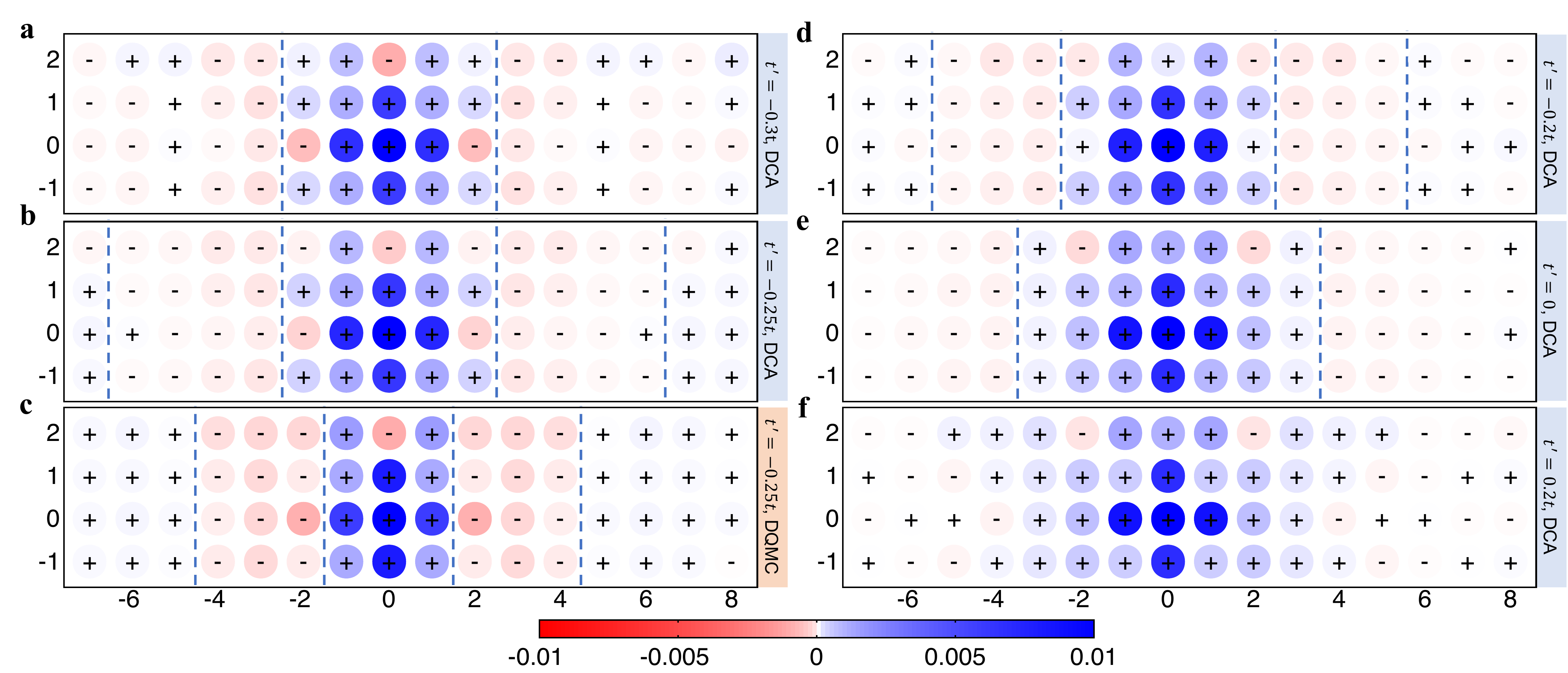}
    \caption{The real-space equal-time staggered spin-spin correlation function of the single-band Hubbard model with $\langle n\rangle=0.8$, obtained from DCA and DQMC simulations. Results are shown for {\bf a} $t^\prime = -0.3t$, DCA; {\bf b} $t^\prime = -0.25t$, DCA; {\bf c} $t^\prime = -0.25t$, DQMC; {\bf d} $t^\prime = -0.2t$, DCA; {\bf e} $t^\prime = 0$, DCA; and {\bf f} $t^\prime = 0.2t$, DCA. The DCA results were obtained using a $16\times4$ cluster embedded in a dynamical mean-field and at an inverse temperature $\beta=6/t$ (Panels {\bf a}-{\bf b} and {\bf d}-{\bf f}). The DQMC results shown in Panel {\bf c} were obtained on a $16\times 4$ cluster with periodic boundary conditions and  $\beta=5/t$ and $t^\prime=-0.25t$. }
    \label{fig:equal_time_spin}
\end{figure*}

\begin{figure}[h!]
    \centering
    \includegraphics[width=\textwidth]{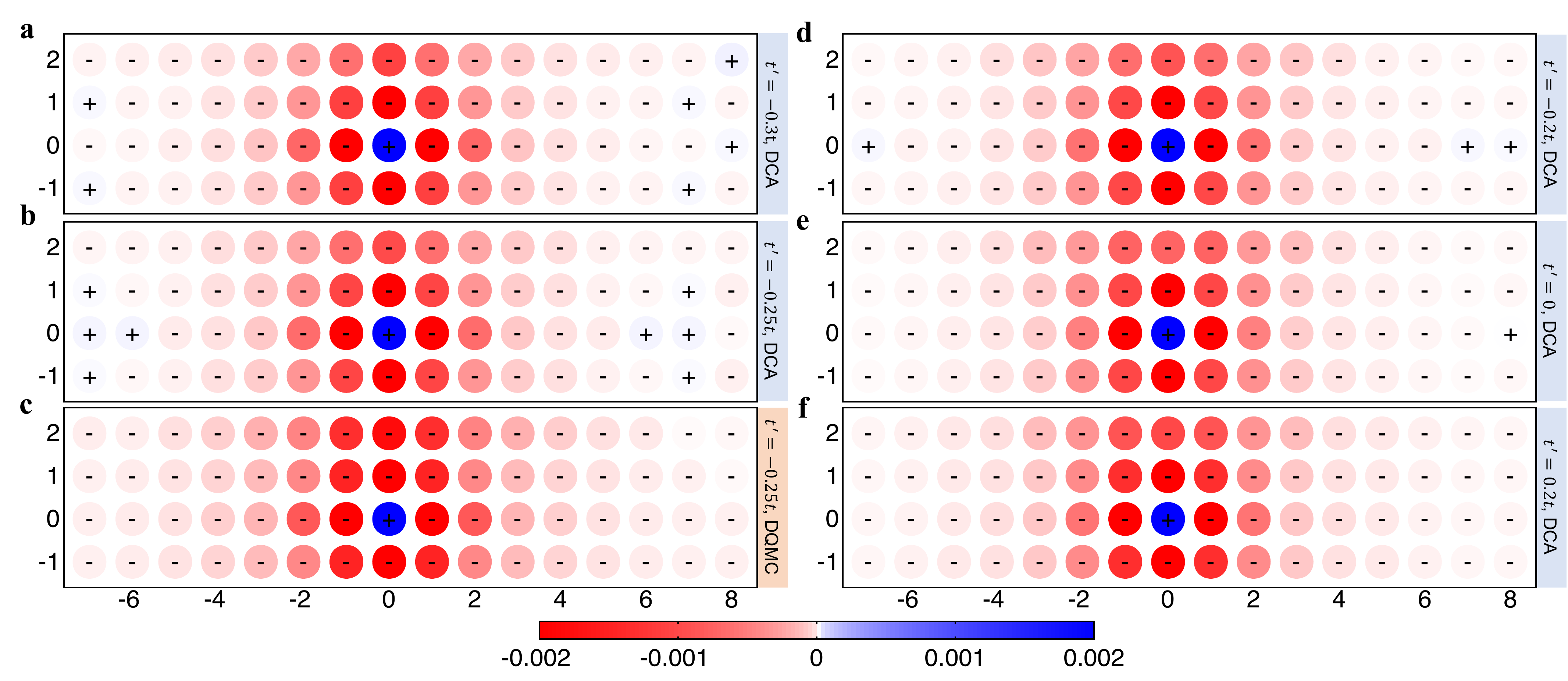}
    \caption{The real-space equal-time density-density correlation function of the single-band Hubbard model with $\langle n\rangle=0.8$, obtained from DCA and DQMC simulations. Results are shown for parameters in a one-to-one correspondence with those shown in Fig.~\ref{fig:equal_time_spin} }
    \label{fig:equal_time_charge}
\end{figure}

In this section, we show the real-space equal-time spin-spin and density-density correlation function in \figdisp{fig:equal_time_spin} and \figdisp{fig:equal_time_charge} respectively in comparison to the static correlation function in Fig.~1 and Fig.~2 of the main text, for the same $16\times 4$ DCA and DQMC simulations. 

In \figdisp{fig:equal_time_spin}, we observe that the equal-time staggered spin-spin correlation function presents a similar pattern to its static counterpart in Fig.~1, with a smaller amplitude. On the other hand, the structure in \figdisp{fig:equal_time_charge} is completely different from that in Fig.~2. For ${\bf r}=0$, $N(0) = n-n^2+2\langle n_\uparrow n_\downarrow\rangle$, so the positive ${\bf r}=0$ correlation is due to the fact that the filling $n=0.8<1$ and the double occupancy $\langle n_\uparrow n_\downarrow\rangle>0$. But the remaining sites are primarily negative for all $t'$, unlike the explicit alternating red and blue regions in Fig.~2, hence showing no signal for charge stripe correlation. This contrast indicates that the imaginary time evolution plays an important role in the formation of the fluctuating charge stripe.

\newpage
\section{The time evolution of the correlation functions}
\begin{figure}[h!]
    \centering
    \includegraphics[width=0.95\textwidth]{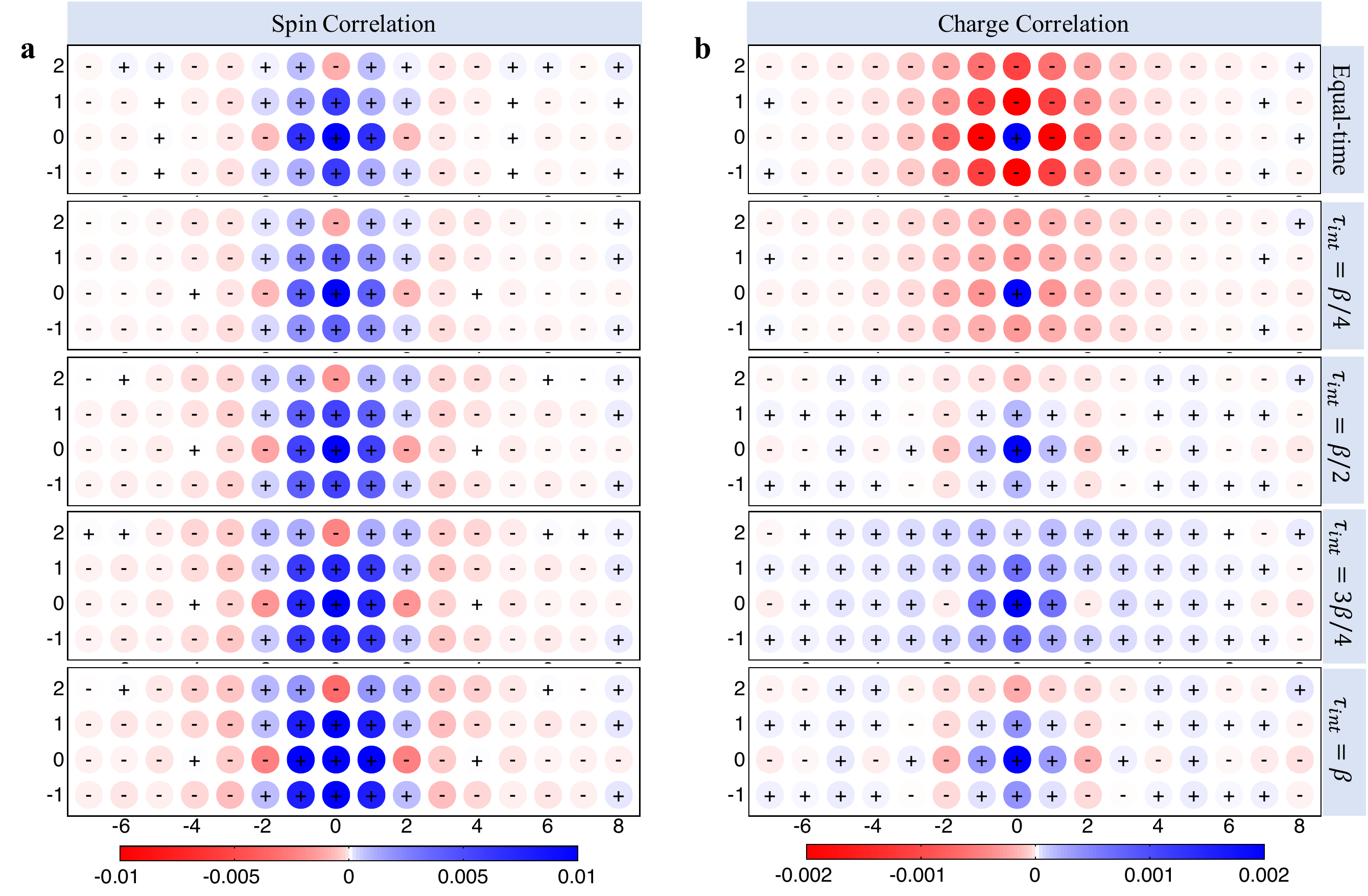}
    \caption{DCA results in real space for integrating the unequal-time spin (panel {\bf a}) and charge (panel {\bf b}) correlation with respect to the imaginary time up to $\tau_{\text{int}}=0,~\beta/4,~\beta/2,~3\beta/4,~\beta$, with the parameter set: $t'=-0.3$, $\langle n\rangle=0.8$, $T=0.167t$ ($\beta=6/t$). }
    \label{fig:DCAtimeintegration1}
\end{figure}

In this section, we analyze how imaginary time integration affects the fluctuating spin and charge stripe pattern. We define 
\begin{equation}
S^\mathrm{stag}({\bf r},\tau_{\text{int}})=(-1)^{r_x+r_y}\tfrac{1}{N}\int_0^{\tau_{\text{int}}} \sum_{\bf i} \langle \hat{S}^z_{\bf i+r}(\tau)~\hat{S}^z_{\bf i}(0)\rangle d\tau
\end{equation} for the spin case, and
\begin{equation}
 N({\bf r},\tau_{\text{int}}) = \frac{1}{N} \int_0^{\tau_{\text{int}}} \sum_{\bf i} \left(\langle n_{{\bf i}+{\bf r}}(\tau)~n_{\bf i}(0)\rangle - \langle n_{{\bf i}+{\bf r}}(\tau)\rangle\langle n_{\bf i}(0)\rangle\right) d\tau   
\end{equation}
for the charge case. We plot the $S^\mathrm{stag}({\bf r},\tau_{\text{int}})$ and $N({\bf r},\tau_{\text{int}})$ for different $\tau_{\text{int}}$ and the equal-time case at $t'=-0.3$, $\langle n\rangle=0.8$, $T=0.167t$ ($\beta=6/t$) in \figdisp{fig:DCAtimeintegration1}. Note that the case with $\tau_{\text{int}}=\beta$ corresponds to the static ($\omega=0$) result in Fig.~1a and Fig.~2a of the main text. This shows how the spin or charge structure evolves from the equal-time to the static case with time integration. In the spin case (\figdisp{fig:DCAtimeintegration1}a), as $\tau_{\text{int}}$ increases, only the amplitude accumulates with almost no change on the distribution of the blue and red region. In contrast, in the charge case (\figdisp{fig:DCAtimeintegration1}b), as $\tau_{\text{int}}$ increases, the pattern starts from almost completely red ($\tau_{\text{int}}=\beta/4$), evolves to almost completely blue ($\tau_{\text{int}}=3\beta/4$), and finally to a stripe-like structure for ($\tau_{\text{int}}=\beta$). This demonstrates that the imaginary time dependence plays an important role in forming the charge stripes.  

\begin{figure}[ht]
    \centering
    \includegraphics[width=0.95\textwidth]{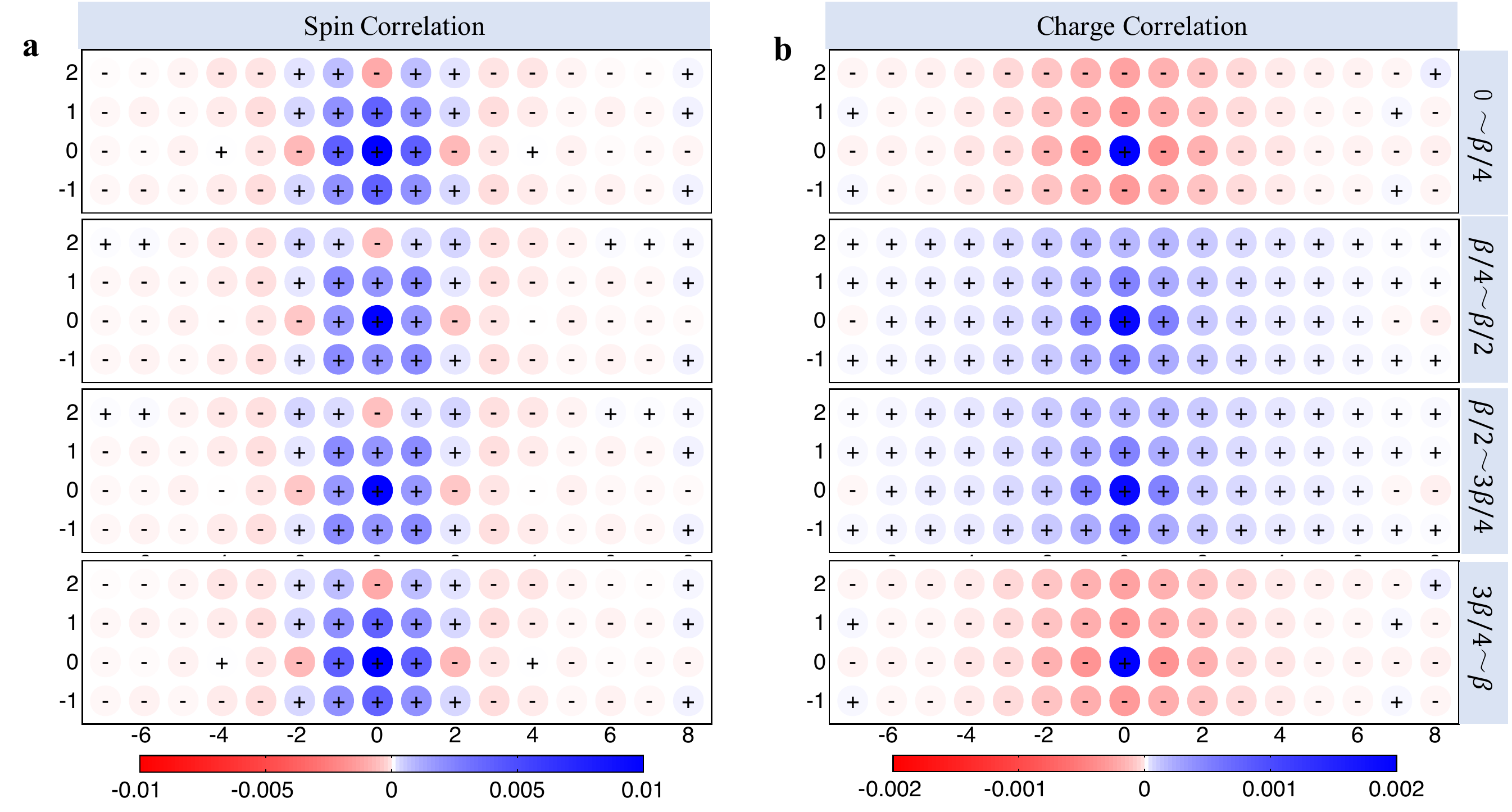}
    \caption{DCA results in real space for integrating the unequal-time spin (panel {\bf a}) and charge (panel {\bf b}) correlation in different imaginary time regions as labeled, with the parameter set: $t'=-0.3$, $\langle n\rangle=0.8$, $T=0.167t$ ($\beta=6/t$).}
    \label{fig:DCAtimeintegration2}
\end{figure}

We can also analyze this result by separating the time integral into quartiles, as shown in \figdisp{fig:DCAtimeintegration2}. Due to the imaginary time symmetry of the correlation functions the results obtained from  integrating on either sides of $\tau=\beta/2$ are identical. The integrated results from $0$ to $\beta/4$ and from $\beta/4$ to $\beta/2$ are quite similar for the spin case, as expected. In the charge case, on the other hand, integrating over $0$ to $\beta/4$ gives an almost completely red region similar to the equal-time case, while the integration over $\beta/4$ to $\beta/2$ shows an almost completely blue region. The resulting charge stripe structure in the static correlation function is a consequence of the contribution from these two patterns.

\clearpage

\section{The effects of different boundary conditions and mean fields on the spin correlations}

\begin{figure}[h!]
    \centering
    \includegraphics[width=\textwidth]{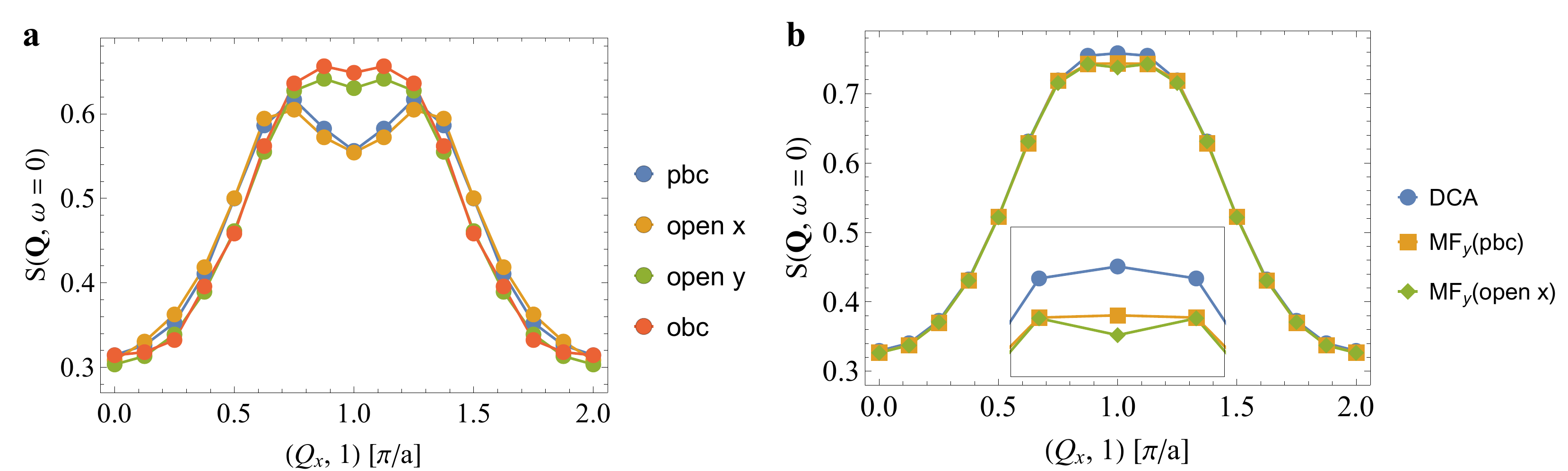}
    \caption{The dynamical spin structure factor at zero frequency $S({\bf Q},\omega = 0)$ for a $16\times4$ cluster with different boundary conditions. Results are shown for a cluster  ({\bf a}) without a mean field and ({\bf b}) embedded in a mean field. Here, ``pbc" means periodic boundary condition in all directions; ``obc" means open boundary condition in all directions; ``open x" means obc along the $x$-axis and pbc along the $y$-axis; likewise, ``open y" means obc along the $y$-axis and pbc along the $x$-axis. Finally, ``MF$_y$" means that we have applied the DCA mean field only along the $y$-axis but not along the $x$-axis. The model parameters in both panels are $t^\prime = -0.3t$, $\langle n \rangle = 0.8$, and $T = 0.22t$ ($\beta=4.5/t$).}
    \label{fig:boundary_effect}
\end{figure}

In this section, we explore the effect of different boundary conditions and the application of different mean fields on the spin correlations. \figdisp{fig:boundary_effect}a presents the dynamical spin structure factor at zero frequency, calculated on a finite-size cluster using continuous-time auxiliary field quantum Monte-Carlo. In this case, we have not coupled the system to a mean-field and applied different boundary conditions to the cluster. Although translational symmetry is broken when the boundary is open in one direction, we assume that the off-diagonal $({\bf k},{\bf k}^\prime)$ elements are negligible and keep only the diagonal $({\bf k},{\bf k})$ elements after Fourier transform. We observe that opening the boundary along the $y$-axis on a $16\times4$ cluster suppresses the spin stripe correlations while opening the boundary along the $x$-axis only has a marginal effect on our results. 

Figure.~\ref{fig:boundary_effect}b explores the effect of the DCA mean-field and the effect of the boundary condition in the presence of the DCA mean-field. Here, a comparison between the blue and orange curves shows that the additional inclusion of the mean field along the $x$-axis in the full DCA result slightly weakens the stripe correlations. Comparing the green and orange curves indicates that opening the left and right boundaries while keeping periodic boundary conditions and the mean-field along the $y$-axis enhances the stripe correlations. 

\begin{figure}[h!]
    \centering
    \includegraphics[width=\textwidth]{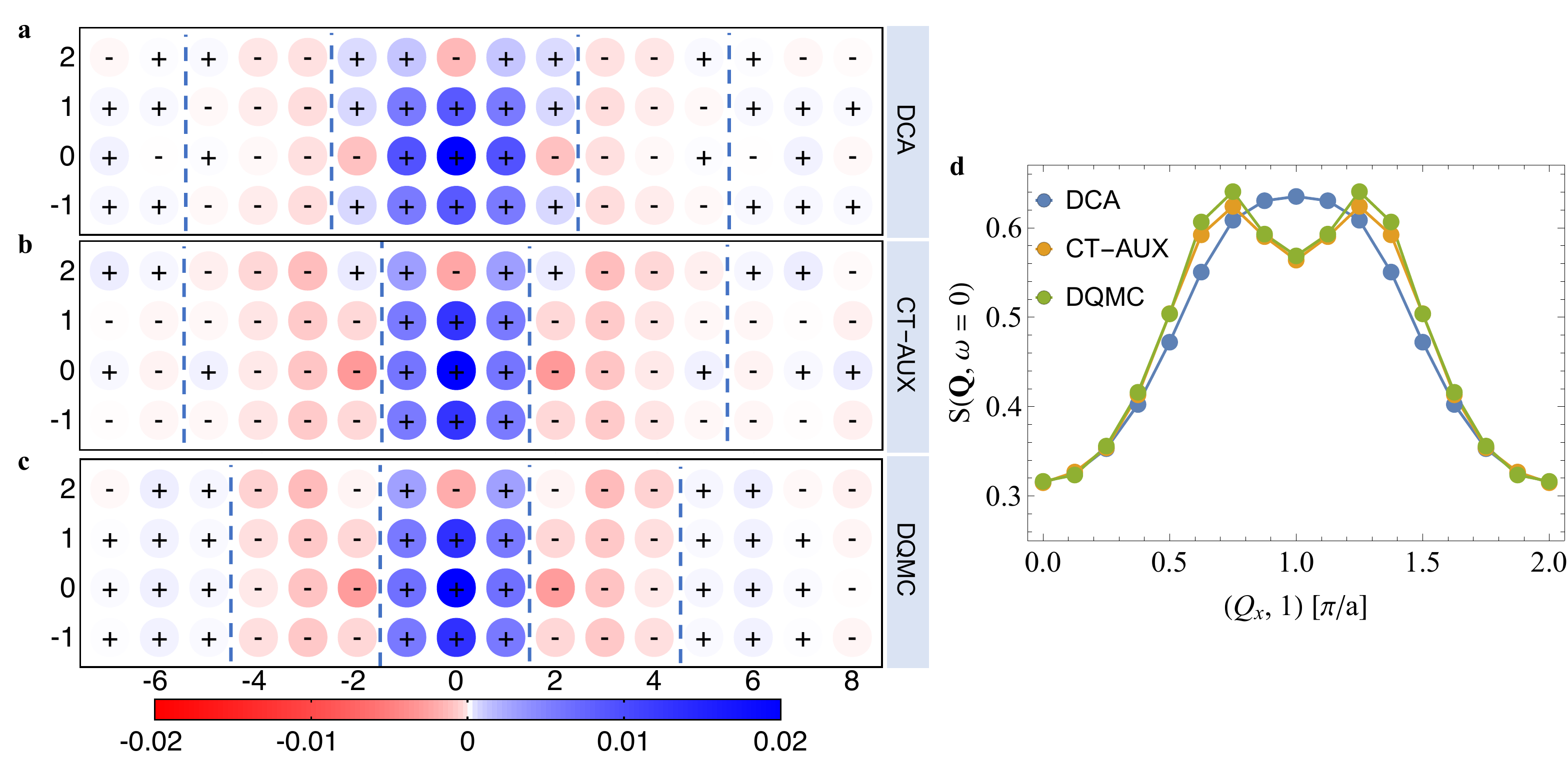}
    \caption{The real-space static staggered spin-spin correlation function in a $16\times4$ cluster ({\bf a}) embedded in a mean field and ({\bf b},{\bf c}) without a mean field. Panel {\bf b} and {\bf c} shows the CT-AUX and DQMC results, respectively. Panel {\bf d} presents the dynamical structure factor for all cases.  The model parameters in both panels are $t^\prime = -0.3t$, $\langle n \rangle = 0.8$, and $T = 0.22t$ ($\beta=4.5/t$).}
    \label{fig:DCA_Stripe_k}
\end{figure}

Figure.~\ref{fig:DCA_Stripe_k} explores the effect of the mean field further. Panels a and (b,c) show  $S^\mathrm{stag}({\bf r})$ for $t^\prime=-0.25t$, $T=0.22t$ ($\beta=4.5/t$) with and without the mean field, respectively, on a cluster with periodic boundary condition. We can see that turning on the mean field weakens the stripe pattern, as indicated by the lighter red region, and expands the central blue region. Panel b and c are from finite-cluster CT-AUX (short for continuous time auxiliary field quantum Monte-carlo, the cluster solver for DCA) and DQMC calculations, respectively. Except for the time-discretization error in DQMC (CT-AUX does not have this error), both methods are numerically exact for a finite cluster. Therefore, the slight difference in the respective results can be attributed to the discrete-time error in DQMC. In panel D, the dip around ${\bf Q} = (\pi,\pi)$ in the finite-cluster CT-AUX and DQMC curves also signals a stronger stripe correlation than the flat peak in the DCA curve.

\section{Systematic comparison between DCA and DQMC results}
\begin{figure}[h!]
    \centering
    \includegraphics[width=\textwidth]{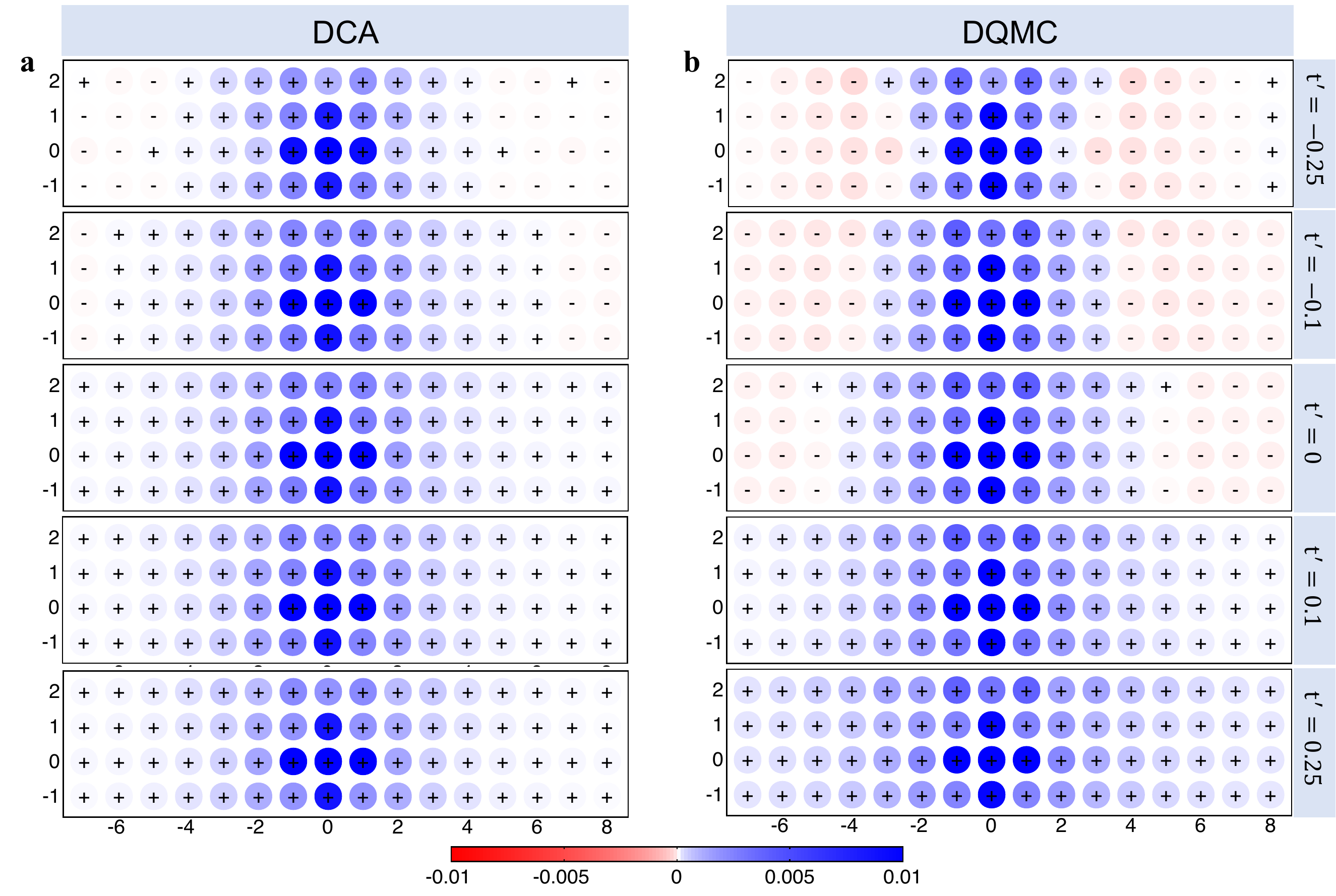}
    \caption{The real-space equal-time staggered spin-spin correlation function in a $16\times4$ cluster calculated using ({\bf a}) DCA and ({\bf b}) DQMC for various values of $t^\prime$, as indicated on the right side of panel ({\bf b}). The model parameters in both panels are $\langle n \rangle = 0.875$, $U = 6t$, and $T = 0.22t$ ($\beta=4.6/t$). 
    Panel ({\bf b}) is adapted from data in Ref.~\cite{HuangQuantMat2018}, which is licensed under \href{https://creativecommons.org/licenses/by/4.0/}{CC BY 4.0}.
    }
    \label{fig:dcavsdqmcn0875}
\end{figure}

Figure \ref{fig:dcavsdqmcn0875} compares the DCA and DQMC (from Ref.~\cite{HuangQuantMat2018}) results for the equal time ($\tau = 0$) staggered spin-spin correlation function for various values of $t^\prime$. Note that our definition of the spin-spin correlation function has an extra factor of $1/4$ compared to that in Ref.~\cite{HuangQuantMat2018}. We therefore divided the DQMC data from Ref.~\cite{HuangQuantMat2018} by a factor of 4. The remaining model parameters are $\langle n \rangle = 0.875$, $U = 6t$, and $T = 0.22t$ ($\beta=4.6/t$), which are identical to the ones used in Ref.~\cite{HuangQuantMat2018}. As with the results at $\langle n \rangle = 0.8$ shown in the main text, we find that DCA predicts weaker spin stripe correlations at this filling $\langle n \rangle = 0.875$ compared to the DQMC results. For the electron-doped cases ($t^\prime>0$), both methods show a full antiferromagnetic (AFM) pattern that extends over the whole cluster. The DQMC signal appears to be stronger (darker blue) especially at longer range. As $t^\prime$ changes from positive to negative, the red regions start to develop for both methods, indicating the appearance of the spin stripe. The DCA stripe is only vaguely seen at this temperature and large negative $t^\prime$, in contrast to a much clearer stripe in DQMC.

\clearpage
\newpage
\section{Temperature dependence of the spin and charge stripes}

\begin{figure}[h!]
    \centering
    \includegraphics[width=\textwidth]{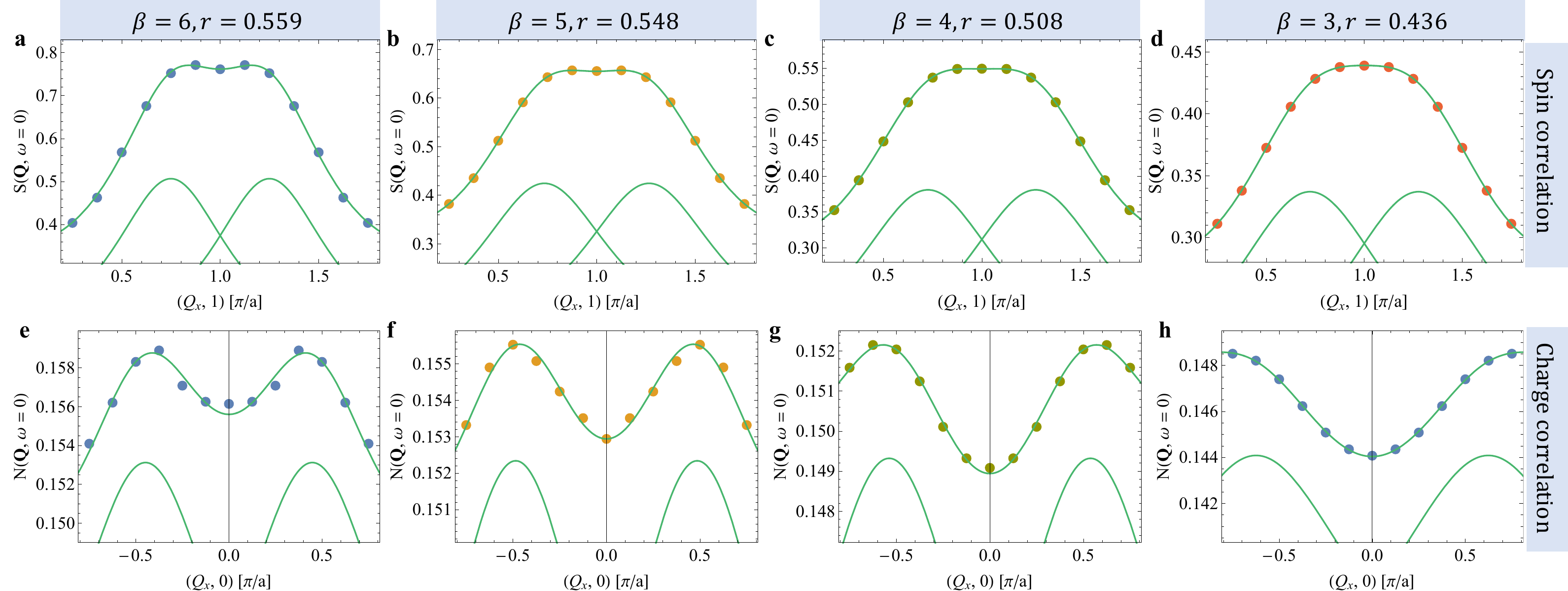}
    \caption{DCA results for the zero-frequency dynamical spin $S({\bf Q},\omega=0)$ (Panels {\bf a-d}) and charge $N({\bf Q},\omega=0)$ (Panels {\bf e-h}) structure factors, obtained on a $16\times4$ cluster embedded in a dynamical mean field. The top row shows the spin structure factors along ${\bf Q} = (Q_x,\pi)$ for {\bf a} $\beta = 6/t$, {\bf b} $\beta = 5/t$, {\bf c} $\beta = 4/t$, and {\bf d} $\beta = 3/t$. Each spectrum is fit with a pair of Lorentzian functions centered at $(2\pi/a)(0.5\pm \delta_s,0.5)$ plus a constant background. The bottom row shows the corresponding charge structure factors along ${\bf Q} = (Q_x,0)$. Each spectrum is fit with a pair of Lorentzian functions centered at ${\bf Q} = (2\pi/a)(\pm\delta_c,0)$ plus a constant background. All results were obtained for $t^\prime = -0.3t$ and $\langle n \rangle = 0.8$. }
    \label{fig:DCA_Stripe_k_weq0}
\end{figure}

To show how the spin and charge stripe form as the system cools down, we plot the zero-frequency dynamical spin and charge structure factors at different temperatures and $t^\prime=-0.3t$ in Fig.~\ref{fig:DCA_Stripe_k_weq0}. As $T$ decreases, $S({\bf Q},\omega = 0)$ (panels a-d) gradually develops two peaks located at $(2\pi/a)(0.5\pm \delta_s,0.5)$. This position is almost independent of temperature. A double peak structure also forms in $N({\bf Q},\omega = 0)$ (panels e-h) as the temperature is lowered, this time centered at $(2\pi/a)(\pm\delta_c,0)$. At high-temperature, $\delta_c = \pi$; however, as the temperature decreases, $\delta_c$ is reduced and the two peaks move towards each other until $\delta_c\approx2\delta_s$ at the lowest temperatures. The measured $\delta_s$ and $\delta_c$ as a function of $T$ are presented in Fig.~3K of the main text.

\clearpage
\newpage
\section{The singlet-pair correlation in a $16\times4$ cluster}
\begin{figure}[h!]
    \centering
    \includegraphics[width=0.55\textwidth]{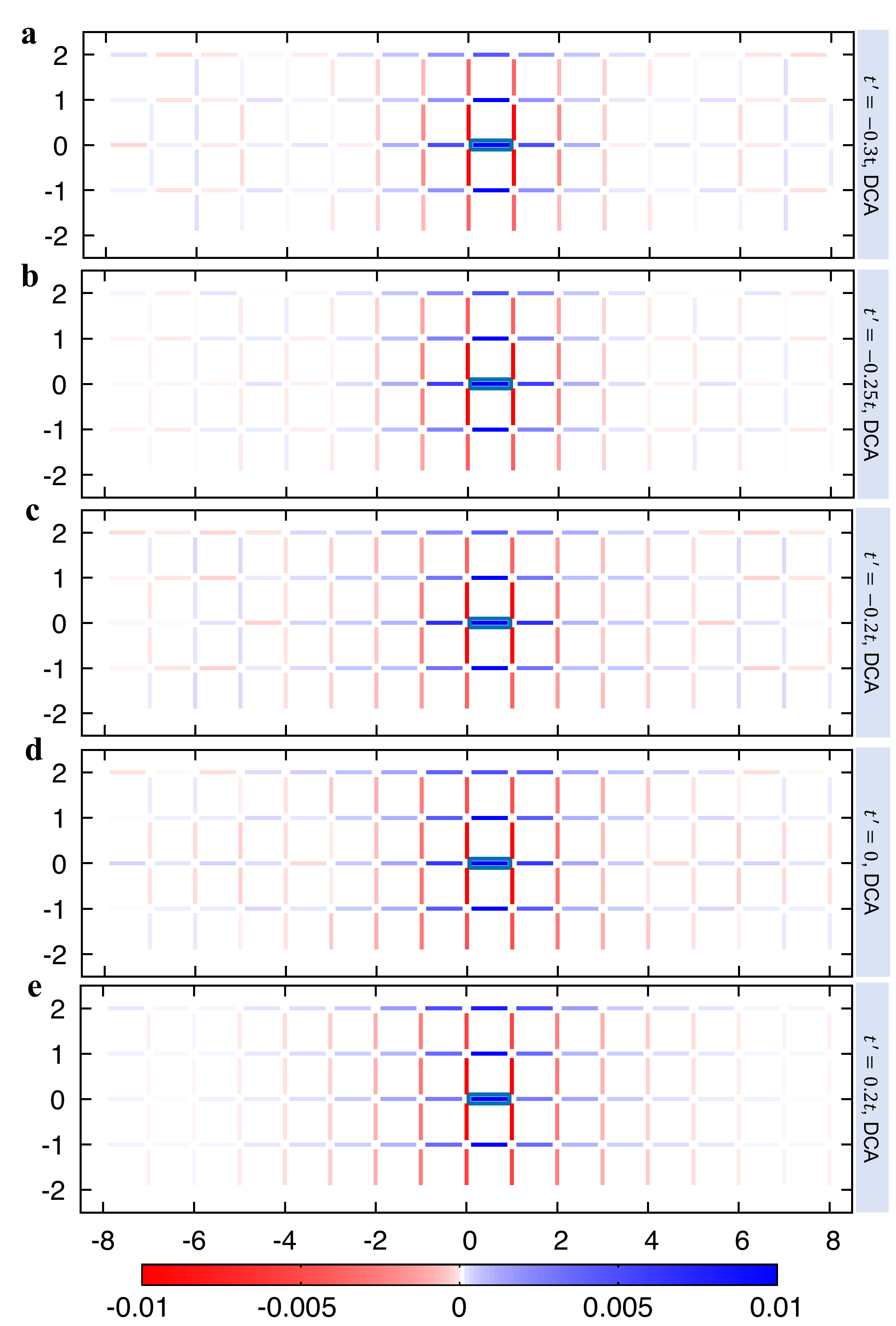}
    \caption{The real-space static singlet-pair correlation functions of the single-band Hubbard model. Results are shown for parameters in one-to-one correspondence with those shown in Fig.~\ref{fig:equal_time_spin}. }
    \label{fig:bonding}
\end{figure}

The singlet-pair correlation function is defined as \cite{Huangjpsp2021}
\begin{equation}
S_{\alpha,\alpha'}({\bf r})=\frac{1}{N}\int_0^\beta d\tau\sum_i\langle \Delta_\alpha(\tau,{\bf r}_i+{\bf r})\Delta^\dagger_{\alpha'}(0,{\bf r}_i) \rangle
\end{equation}
\begin{equation}
\Delta^\dagger_{\alpha'}({\bf r}_i)=\frac{1}{\sqrt{2}}(c^\dagger_{{\bf r}_i,\uparrow}c^\dagger_{{\bf r}_i+\alpha,\downarrow}-c^\dagger_{{\bf r}_i,\downarrow}c^\dagger_{{\bf r}_i+\alpha,\uparrow})
\end{equation}
where $\alpha,~\alpha^\prime=\hat{x}~\text{or}~\hat{y}$. We present this correlation function in \figdisp{fig:bonding} using $S_{\hat{x},\hat{x}}({\bf r})$ and $S_{\hat{y},\hat{x}}({\bf r})$ for the horizontal and vertical bars, respectively. For $t^\prime\ge 0$, the horizontal (vertical) bars are always blue (red), showing a uniform $d$-wave pattern. As $t^\prime$ becomes negative, blue vertical and red horizontal bars start to appear at large distances away from the central site, indicating a $\pi$-phase shift in the singlet correlations consistent with a pair-density-wave (PDW). The results for $t^\prime=-0.25t$ do not follow this trend, possibly due a proximity to the Lifshitz transition, where the Fermi surface changes from hole- to electron-like. These results can be compared with a recent DQMC study on a $8\times8$ cluster \cite{Huangjpsp2021}, where no evidence of a PDW-like pattern was observed. From \figdisp{fig:bonding}{\bf a} and {\bf c}, one sees that the PDW-like region appears at $|r_x|\ge5$, outside the range $|r_x|\le4$ of the $8\times8$ cluster. Therefore, our results are consistent with the DQMC results on the $8\times8$ cluster. 

\newpage
\section{Our Customized Color Bar}
The color bars we used throughout this paper are derived from a customized color map defined to highlight the differences in the correlations at small values. 
Our definition for this map is the same as the one used in Refs.~\cite{HuangQuantMat2018} and \cite{HuangScience2017}. It is defined using the RGB (Red, Green, Blue) scheme. First, we define a piecewise function: $f(x)=1$ if $x < 0.5$; $f(x)=1-\sqrt{2x-1}$ if $x > 0.5$, setting $x$ in the range $[0,1]$. Then
\begin{equation}
\begin{split}
&r = f(x), \\&
g = \mbox{min}[f(x),f(1-x)],\\&
b = f(1-x),
\end{split}
\end{equation}
where $r$, $g$, $b$ stand for red, green, blue color, respectively.

\bibliography{refbib}

\FloatBarrier